\documentclass[12pt]{article}

\usepackage{putex}
\usepackage{amsmath}
\usepackage{amssymb}
\usepackage{cite}
\usepackage{bigfoot}
\usepackage[
      colorlinks=true,
      linkcolor=blue,
      urlcolor=blue,
      filecolor=black,
      citecolor=red,
      linktocpage=true,
      bookmarksdepth=2
      ]{hyperref}
\hypersetup{pdftitle={Reduced correlators and one-loop scattering in AdS},pdfauthor={Shai M. Chester, Rishi Mouland, Cl\'ement Virally}}

\newcommand {\be} {\begin {equation}}
\newcommand {\ee} {\end {equation}}
\newcommand {\nn} {\nonumber}

\newcommand {\bes} {\begin {equation*}}
\newcommand {\ees} {\end {equation*}}

\newcommand{\es}[2]{%
  \begin{equation}
    \begin{aligned}
      #2
    \end{aligned}
    \phantomsection\label{#1}%
  \end{equation}%
}

\newcommand{\Z}{\mathbb{Z}}
\newcommand{\N}{\mathbb{N}}
\newcommand{\R}{\mathbb{R}}

\newcommand{\cA}{{\mathcal A}}
\newcommand{\cB}{{\mathcal B}}

\newcommand{\cD}{{\mathcal D}}

\newcommand{\cG}{{\mathcal G}}

\newcommand{\cL}{{\mathcal L}}
\newcommand{\cN}{{\mathcal N}}
\newcommand{\cO}{{\mathcal O}}

\newcommand{\cM}{{\mathcal M}}

\renewcommand{\r}{{\mathbf{r}}}

\renewcommand{\d}{{\mathbf{d}}}

\newcommand{\bea}{\begin{equation}\begin{aligned}}
\newcommand{\eea}[1]{\label{#1}\end{aligned}\end{equation}}

\newcommand{\beq}{\begin{equation}}
\newcommand{\eeq}{\end{equation}}

\def\ie{\begin{equation}\begin{aligned}}
\def\fe{\end{aligned}\end{equation}}

\numberwithin{equation}{section}

\def\<{\langle}
\def\>{\rangle}

\newcommand{\h}{h^\vee}

\begin{document}

\preprint{Imperial--TP--2026--SMC--2}

\institution{imperial}{Abdus Salam Centre for Theoretical Physics, Imperial College London,\cr London SW7 2AZ, UK}
\institution{oxford}{Mathematical Institute, University of Oxford,
Woodstock Road, Oxford, OX2 6GG, UK }

\title{Reduced correlators and one-loop scattering in AdS}

\authors{Shai M.~Chester,\worksat{\imperial} Rishi Mouland,\worksat{\imperial} and Cl\'ement Virally\worksat{\oxford}}

\abstract{
We consider four-point functions of half-BPS operators in AdS$_{d+1}$/CFT$_d$ dual to scattering of gravitons in maximally supersymmetric theories and gluons in half-maximal supersymmetric theories. In even $d=4,6$, the superconformal Ward identities were previously solved by expressing the correlator as a differential operator acting on a reduced correlator, but in odd dimensions this construction becomes non-local due to fractional derivatives. We introduce a new local reduced correlator in position space that is well defined in all dimensions allowed by supersymmetry: $d=3,4,6$ for maximal and $d=3,4,5,6$ for half-maximal. The reduced correlator obeys simple crossing equations, admits a superblock expansion in terms of a single block, and for $d=3$ makes manifest the protected 1d topological subsector. Using reduced correlators, we compute one-loop gluon scattering for $d=3,5,6$, generalizing the known $d=4$ result, and simplify previous one-loop graviton scattering results in $d=3,6$. We match the gluon correlators to the flat space limit in all dimensions and, in $d=3$, to protected OPE coefficients previously obtained from supersymmetric localization.
}
\date{}

\hypersetup{pageanchor=false}
\maketitle
\hypersetup{pageanchor=true}

\tableofcontents
\newpage

\section{Introduction}\label{sec:intro}

Four-point functions of half-BPS operators in supersymmetric conformal field theories (CFTs) are fundamental observables in AdS/CFT. For examples of AdS/CFT with maximal supersymmetry, which exist for CFTs in spacetime dimensions $d=3,4,6$, the Kaluza-Klein (KK) mode of the graviton labeled by $p=2,3,\dots$ is dual to a half-BPS multiplet whose scalar superprimary has $\Delta=\frac{d-2}{2}p$. For instance, for $p=2$ this is the stress tensor multiplet, which is dual to the graviton multiplet in AdS$_{d+1}$ \cite{Maldacena:1997re}. One can also consider examples of AdS/CFT with half-maximal supersymmetry that have flavor branes in the bulk, with a fixed point locus AdS$_{d+1}\times S^3$. The KK mode labeled by $p$ from gluons on the brane is then dual to a half-BPS multiplet whose scalar superprimary again has $\Delta=\frac{d-2}{2}p$, where $p=2$ now corresponds to the flavor multiplet, which is dual to the gauge theory on the brane \cite{Aharony:1998xz,Alday:2021odx}.\footnote{For half-maximal supersymmetry there are now several towers of graviton modes, most of which are not half-BPS.} Four-point functions of these half-BPS operators are then dual to scattering of the dual graviton or gluon modes in the bulk. In 4d, the correlators have been studied in a weak coupling expansion \cite{Eden:1998hh,Eden:2000mv,Eden:2000bk,Arutyunov:2003ad,Chicherin:2014esa,Chicherin:2015edu,Chicherin:2018avq,Du:2024xbd}. For general $d$, they have been studied at strong coupling using the analytic bootstrap \cite{Alday:2020tgi,Alday:2021ymb,Alday:2020dtb,Alday:2022uxp,Alday:2022xwz,Alday:2020lbp,Alday:2018pdi,Alday:2021odx,Alday:2017xua,Aharony:2016dwx,Alday:2021vfb,Alday:2022rly,Alday:2018kkw,Alday:2021ajh,Alday:2019nin,Alday:2024yax,Huang:2023oxf,Rastelli:2017ymc,Rastelli:2017udc,Zhou:2018ofp,Zhou:2017zaw,Huang:2023ppy,Binder:2019jwn,Binder:2018yvd,Binder:2019mpb,Chester:2023qwo,Chester:2025jxg,Chester:2020dja,Chester:2019pvm,Chester:2025ssu,Chester:2018dga,Chester:2020vyz,Chester:2018aca,Behan:2024vwg,Chester:2025wti,Caron-Huot:2018kta,Alday:2017vkk}, and even at finite coupling using the numerical bootstrap \cite{Chester:2023ehi,Chester:2021aun,Beem:2016wfs,Alday:2022ldo,Beem:2013qxa,Beem:2015aoa,Beem:2014zpa,Chester:2022sqb,Agmon:2019imm,Chester:2014mea,Chester:2014fya,Chester:2024bij,Chang:2017cdx,Chang:2019dzt,Chang:2017xmr,Chester:2026uro,Caron-Huot:2024tzr,Agmon:2017xes}.

The constraints of superconformal symmetry on these correlators are imposed by differential equations called the superconformal Ward identities. They take the remarkably simple form \cite{Dolan:2004mu}
\es{wardGen}{
\Big(z\partial_z-\frac{d-2}{2}\alpha\partial_\alpha\Big)\mathcal{G}(z,\bar z,\alpha,\bar\alpha)\vert_{\alpha=1/z}=0\,,
}
as well as three additional equations obtained from $z\leftrightarrow\bar z$ and $\alpha\leftrightarrow\bar\alpha$ for maximal supersymmetry, and one additional equation from $z\leftrightarrow\bar z$ for half-maximal supersymmetry. Here, $z,\bar z$ and $\alpha,\bar\alpha$ are conformal and R-symmetry cross-ratios, respectively, $\mathcal{G}(z,\bar z,\alpha,\bar\alpha)$ is the four-point function with an overall kinematic factor stripped off, and the half-maximal correlator does not depend on $\bar\alpha$ (since the operators we consider are charged under only one $SU(2)$ factor of the R-symmetry). 

For $p=2$, the Ward identities have a formal solution for general $d$ for maximal and half-maximal supersymmetry \cite{Dolan:2004mu}:
\es{wardSolForm}{
\mathcal{G}(z,\bar z,\alpha,\bar\alpha)&=(z\bar z)^{d-2}\nabla_d^{\frac{d-4}{2}}(z\bar z)^{\frac{d-4}{2}}[(1-z\alpha)(1-\bar z\alpha)(1- z\bar\alpha)(1-\bar z\bar\alpha)]\overline{\mathcal{G}}(z,\bar z)\,,\\
\cG(z,\bar z,\alpha)&=(z\bar z)^{d-2}\nabla_d^{\frac{d-4}{2}}(z\bar z)^{\frac{d-4}{2}}[(1-z\alpha)(1-\bar z\alpha)]\overline{\cG}(z,\bar z)\,,\\
}
where $\nabla_d=\partial_z\partial_{\bar z}-\frac{d-2}{2(z-\bar z)}(\partial_z-\partial_{\bar z})$ is proportional to the Laplacian in $d$ dimensions, and the so-called reduced correlator $\overline{\mathcal{G}}$ is (two) four degrees smaller in $\alpha,\bar\alpha$ for (half-)maximal supersymmetry. The reduced correlator transforms simply under crossing symmetry. It also has a very simple expansion in superblocks \cite{Beem:2013qxa,Beem:2014zpa,Beem:2015aoa}, where just a single block appears for each supermultiplet, unlike the full correlator where superblocks are complicated sums of conformal blocks for each operator in the multiplet. In 6d for maximal supersymmetry or 4d for both half-maximal and maximal supersymmetry, protected multiplets are described by a 2d chiral algebra \cite{Beem:2013sza,Beem:2014kka}, which can be obtained from the full correlator by performing the twist $\alpha=1/\bar z$ (in 6d, $\alpha=\bar\alpha=1/\bar z$). The reduced blocks for these protected multiplets are singular under this twist, unless we include another term in the solution to the Ward identities that depends on a holomorphic function $f(z)$. Upon performing the twist, we then find that the contribution of the reduced correlator vanishes, and the full correlator reduces to $f(z)$ (which in 4d for maximal supersymmetry also depends on $\bar\alpha$), as we expect for a 2d chiral algebra \cite{Beem:2013sza,Beem:2014kka}. This holomorphic function obeys its own separate crossing equation, which is equivalent to crossing in 2d.

The problem with the formal solution in \eqref{wardSolForm} is that it is only well defined for $d=4,6$, as for odd $d$ one has fractional derivatives, which makes the map between full and reduced correlators non-local. While one can still formally expand the reduced correlator in blocks for odd $d$ at least for a subset of half-BPS correlators \cite{Woolley:2026cii,Woolley:2026anb}, this makes the physical meaning of the reduced correlator unclear. Another issue is that for 3d, the theory has a 1d topological subsector that describes protected operators \cite{Chester:2014mea}, analogous to the 2d chiral algebra for 4d and 6d, which can be obtained from the topological twist $z=\bar z=1/\alpha$. We would like a formal solution to the Ward identity that makes this 1d sector manifest, just as the 2d chiral algebra was manifest as described above.

A solution to the problem of the fractional derivative was recently found in Mellin space for maximal \cite{Virally:2025nnl} and half-maximal supersymmetry \cite{Chester:2025jxg}. In the following we will focus on the correlator of $p=2$ multiplets for simplicity. The Mellin transform of the connected part of the full correlator is defined as
\es{mellinDef}{
\cG(U,V;\alpha,\bar\alpha)&=\int\frac{\text{d}s\,\text{d}t}{(4\pi i)^2}U^{\frac s2}V^{\frac{t}{2}-d+2}\Gamma^2\Big[d-2-\frac s2\Big]\Gamma^2\Big[d-2-\frac t2\Big]\\
&\quad\times\Gamma^2\Big[d-2-\frac u2\Big]\mathcal{M}(s,t;\alpha,\bar\alpha)\,,
}
where $U=z\bar z$, $V=(1-z)(1-\bar z)$, $u=4(d-2)-s-t$, the $s,t$ contours are chosen such that all poles of each gamma function lie on the same side of the contour, and for half-maximal supersymmetry we simply drop the $\bar\alpha$. Derivatives in $U,V$ (or similarly $z,\bar z$) become simple multiples of $s,t$ in Mellin space, which makes it plausible that a reduced correlator in Mellin space can be defined using a local map for all $d$, as was accomplished in \cite{Virally:2025nnl,Chester:2025jxg}. Furthermore, the Mellin amplitude can be expanded in exchange Witten diagrams, which are the Mellin space analogue of conformal blocks. This reduced Mellin amplitude was useful for simplifying analytic bootstrap calculations. The drawback of Mellin space is that there is no simple expression for conformal blocks in Mellin space, and certain terms in the full correlator, such as the rational terms associated with the protected subsectors, have no Mellin amplitude. So for the numerical bootstrap, it is still desirable to have a reduced correlator in position space.

In this paper, starting from the reduced Mellin amplitude we derive a reduced correlator in position space that is local for all $d$. For 3d maximal supersymmetry, the map that relates the full to the reduced correlator has up to four derivatives in $U,V$, while for half-maximal supersymmetry the map has up to two derivatives in $U,V$. Just as in the previously known 4d and 6d cases, the reduced correlator obeys simple crossing equations, up to elements of the kernel of the map to the full correlator, and has a superblock expansion with just one block per supermultiplet. In 3d, we find that the reduced superblocks for protected multiplets are singular under the twist $z=\bar z=1/\alpha$ unless we include extra terms in the solution to the Ward identities, which are just polynomials in $U,V,\alpha,\bar\alpha$. Upon performing the twist, we then find that the contribution of the reduced correlator vanishes, and the extra terms lose their dependence on $U,V$, as we expect for a 1d topological correlator. Note that in 5d no such extra terms are needed, which is consistent with the absence of a protected subsector there. 

We then use this correlator to compute the strong coupling expansion of the gluon correlator to one loop for half-maximal supersymmetry in $d=3,5,6$, generalizing the previous answer for 4d \cite{Alday:2021ajh}. We check that the flat space limit of our AdS$_{d+1}\times S^3$ holographic correlator matches the expected flat space amplitude, which is a sum of scalar box diagrams in $d+4$ dimensions with the correct coefficient. We use the Lorentzian inversion formula \cite{Caron-Huot:2017vep} to extract CFT data in all cases, which in 3d we match to the protected OPE coefficients that follow from supersymmetric localization \cite{Chester:2023qwo,Kubo:2025dot}. We also use our reduced correlators to simplify the previously known one-loop corrections to graviton scattering in $d=3,6$ with maximal supersymmetry \cite{Alday:2020tgi,Alday:2021ymb,Alday:2022rly}.

The rest of this paper is organized as follows. In Section \ref{sec:corrMax} we consider the stress tensor correlator for 3d maximal supersymmetry, and introduce our new reduced correlator in that case. In Section \ref{sec:corrHalf}, we consider the flavor multiplet correlator for half-maximal supersymmetry in $d=3,5,6$ and introduce our new reduced correlator. In Section \ref{sec:1loop}, we use these reduced correlators to compute the one-loop correction to gluon scattering, extract CFT data, compare to localization for 3d, and check the flat space limit. We conclude in Section \ref{sec:conclusion} with a review of our results and a discussion of future directions. Technical details of the calculations are given in the various Appendices, including a simplified version of the one-loop graviton scattering results from \cite{Alday:2020tgi,Alday:2021ymb,Alday:2022rly}.

\section{Stress tensor correlator for 3d \texorpdfstring{$\mathcal{N}=8$}{N=8}}\label{sec:corrMax}
	\subsection{Full correlator}
	\label{sec:fullMax}

    In maximally supersymmetric superconformal field theories (SCFTs), we study correlation functions of half-BPS scalars $\mathcal{O}_{I_1,\ldots,I_p}(x)$, which transform under the rank-$p$ symmetric traceless representation of the R-symmetry group $SO(n)$, with $n=8,6,5$ for $d=3,4,6$. These have protected conformal dimensions $\Delta = \frac{d-2}{2}p$, where $p\ge 2$ is an integer. It is convenient to contract their indices with auxiliary null polarization vectors $t^I$, $t^I t_I=0$, so that:
    \begin{equation}
        \mathcal{O}_p(x,t) \equiv t^{I_1}\ldots t^{I_p}\mathcal{O}_{I_1,\ldots,I_p}(x)\;.
    \end{equation}
    We will be interested in four-point functions of identical operators with $p=2$, denoted $\langle 2222\rangle$. These are fixed by conformal and $SO(n)$ symmetry to take the form:
    \begin{equation}
        \langle\mathcal{O}_2(x_1,t_1)\mathcal{O}_2(x_2,t_2)
        \mathcal{O}_2(x_3,t_3)\mathcal{O}_2(x_4,t_4)\rangle = \frac{(t_1\cdot t_2)^2(t_3\cdot t_4)^2}{|x_{12}|^{2(d-2)}|x_{34}|^{2(d-2)}}\mathcal{G}(U,V;\sigma,\tau)\;,
    \end{equation}
    where $U,V$ are the conformal cross-ratios and $\sigma,\tau$ are the R-symmetry cross-ratios:
    \begin{equation}
        U = \frac{x_{12}^2x_{34}^2}{x_{13}^2x_{24}^2}\;,\quad V = \frac{x_{14}^2x_{23}^2}{x_{13}^2x_{24}^2}\;,\quad \sigma = \frac{(t_1\cdot t_3)(t_2\cdot t_4)}{(t_1\cdot t_2)(t_3\cdot t_4)}\;,\quad \tau = \frac{(t_1\cdot t_4)(t_2\cdot t_3)}{(t_1\cdot t_2)(t_3\cdot t_4)}\;,\label{eq:cross_ratios_max}
    \end{equation}
    with $x_{ij}=x_i-x_j$. Note that, since the correlator is quadratic in each of the $t_i$, it is quadratic in $\sigma,\tau$. These are the bosonic constraints of the superconformal group. Invariance under the action of the fermionic generators further imposes the superconformal Ward identities (SCWI), which take the universal form (\ref{wardGen}).
    The cross-ratios $z,\bar z,\alpha,\bar\alpha$ are related to $U,V,\sigma,\tau$ by:
    \begin{equation}
        U = z\bar z\;,\quad V=(1-z)(1-\bar z)\;,\quad \sigma=\alpha\bar\alpha\;,\quad \tau=(1-\alpha)(1-\bar\alpha)\;.
    \end{equation}
    To solve these constraints, we can expand the correlator in superconformal blocks:
    \begin{equation}
        \mathcal{G}(U,V;\sigma,\tau) = \sum_{\chi}\lambda^2_\chi\mathcal{G}_\chi(U,V;\sigma,\tau)\;,
    \end{equation}
    where $\chi$ runs over all multiplets in the $\mathcal{O}_2\times\mathcal{O}_2$ OPE, and each individual superconformal block $\mathcal{G}_\chi(U,V;\sigma,\tau)$ solves the SCWI. The superconformal blocks themselves can be written in terms of the standard bosonic conformal blocks $G_{\Delta,\ell}(U,V)$:
    \begin{equation}
        \mathcal{G}_{\chi}(U,V;\sigma,\tau) = \sum_{0\leq j\leq i\leq 2}\sum_{\Delta',\ell'} f^{i,j}_{\Delta',\ell'}Y_{ij}(\sigma,\tau)G_{\Delta',\ell'}(U,V)\;.
    \end{equation}
    The coefficients $f^{i,j}_{\Delta',\ell'}$ can be obtained by demanding that this should solve the SCWI, which is a tedious but straightforward exercise. The exact forms of the polynomials $Y_{ij}(\sigma,\tau)$, as well as the coefficients $f^{i,j}_{\Delta',\ell'}$ for each of the exchanged multiplets, can be found in~\cite{Chester:2014fya} for the 3d $\mathcal{N}=8$ theory. In terms of the blocks defined there, the full correlator in this theory can therefore be written:
    \begin{equation}
        \begin{aligned}
            \mathcal{G}(U,V;\sigma,\tau) ={}& \lambda^2_1 + \lambda^2_{(B,2)}\mathcal{G}_{(B,2)}(U,V;\sigma,\tau) + \lambda^2_{(B,+)}\mathcal{G}_{(B,+)}(U,V;\sigma,\tau)\\
            &+\lambda^2_{\text{Stress}}\mathcal{G}_{\text{Stress}}(U,V;\sigma,\tau)+\sum_{\ell\ \text{odd}} \lambda^2_{(A,2)_\ell}\mathcal{G}_{(A,2)_\ell}(U,V;\sigma,\tau)\\
            &+\sum_{\ell\ \text{even}} \lambda^2_{(A,+)_\ell}\mathcal{G}_{(A,+)_\ell}(U,V;\sigma,\tau)+\sum_{\ell\ \text{even}}\,\sum_{\Delta>\ell+1}\lambda^2_{\Delta,\ell}\mathcal{G}_{\Delta,\ell}(U,V;\sigma,\tau)\;,
        \end{aligned}
    \end{equation}
    where the last sum is over the long multiplets and $\lambda^2_1$ is the coefficient of the identity, which equals $1$ for unit-normalized $\mathcal{O}_2$. The $\langle 2222\rangle$ correlator is symmetric under the permutation of any of the external operators, which is known as crossing symmetry, and imposes the crossing relations:
    \begin{equation}
        \mathcal{G}(U,V;\sigma,\tau) = \mathcal{G}\left(\frac{U}{V},\frac{1}{V};\tau,\sigma\right) =U^{d-2}\sigma^2\mathcal{G}\left(\frac{1}{U},\frac{V}{U};\frac{1}{\sigma},\frac{\tau}{\sigma}\right)  = \frac{U^{d-2}}{V^{d-2}}\tau^2\mathcal{G}\left(V,U;\frac{\sigma}{\tau},\frac{1}{\tau}\right)\;.\label{eq:crossing_pos_max}
    \end{equation}
    In 3d $\mathcal{N}=8$, there is an interesting kinematic point to study. We consider the twisted correlator, which is to say $\mathcal{G}(U,V;\sigma,\tau)$ evaluated at the point $\bar z = z = \frac{1}{\alpha}$. As a consequence of the SCWI, we can show that:
    \begin{equation}
        \frac{\text{d}}{\text{d}z}\Big[\mathcal{G}(U,V;\sigma,\tau)\Big|_{\bar z=z,\ \alpha=\frac{1}{z}}\Big]=\Big(\partial_z+\partial_{\bar z}-\alpha^2\partial_\alpha\Big)\mathcal{G}(U,V;\sigma,\tau)\Big|_{\bar z=z = \frac{1}{\alpha}}=0\;,
    \end{equation}
    such that after twisting, the correlator is a function of $\bar\alpha$ alone on each of the intervals $z<0$, $0<z<1$ and $z>1$. It is simple to show that, after performing the twisting, all superconformal blocks vanish, except for those of the identity, $B$-type and stress tensor multiplets. Defining the Legendre polynomial $P_j(x)$, the twisted correlator for $0<z<1$ is explicitly given by
    \begin{equation}
        \mathcal{G}(U,V;\sigma,\tau)\Big|_{\bar z=z = \frac{1}{\alpha}}=\lambda_1^2 + \frac{\lambda_{(B,2)}^2}{16} + \frac{\lambda_{(B,+)}^2}{16}P_2(2\bar\alpha-1) + \frac{\lambda_{\text{Stress}}^2}{4}P_1(2\bar\alpha-1)\;.\label{eq:max_susy_twisted}
    \end{equation}
    Thus, under crossing, we can obtain a relationship between the OPE coefficients of these blocks. If we use the last relation in (\ref{eq:crossing_pos_max}), which maps the interval $0<z<1$ to itself, and perform the twisting on both sides, we obtain the twisted crossing equation for 3d $\mathcal{N}=8$~\cite{Chester:2014mea}:
    \begin{equation}
        4\lambda^2_{\text{Stress}}-5\lambda^2_{(B,+)}+\lambda_{(B,2)}^2+16\lambda_1^2=0\;.\label{eq:crossing_twisted_max}
    \end{equation}
    Often, it will be convenient to represent the correlators in Mellin space. The Mellin amplitude $\mathcal{M}(s,t;\sigma,\tau)$, which encodes the dynamical information of $\mathcal{G}(U,V;\sigma,\tau)$, is defined by the integral transform~(\ref{mellinDef}). In Section~\ref{sec:redMax}, we will show that the Mellin transform provides an alternative way to solve the SCWI. In Mellin space, the crossing equation~(\ref{eq:crossing_pos_max}) becomes:
    \begin{equation}
        \mathcal{M}(s,t;\sigma,\tau) = \mathcal{M}(s,u;\tau,\sigma)=\sigma^2\mathcal{M}\left(u,t;\frac{1}{\sigma},\frac{\tau}{\sigma}\right) = \tau^2\mathcal{M}\left(t,s;\frac{\sigma}{\tau},\frac{1}{\tau}\right)\;.\label{eq:mellin_crossing_max}
    \end{equation}
	
	\subsection{Reduced correlator}\label{sec:redMax}
    There is another approach to solving the SCWI which, in the simplest cases, avoids carrying around  the R-symmetry cross-ratios. In~\cite{Dolan:2004mu}, it was shown that there exists a differential operator which generates solutions to the SCWI, i.e.\ solutions can be shown to take the form~(\ref{wardSolForm}). In these solutions, the reduced correlator $\overline{\mathcal{G}}(z,\bar z)$ contains all of the dynamical information, while the $\nabla_d$ operators are defined such that, for any choice of the reduced correlator, the full correlator $\mathcal{G}(z,\bar z;\alpha,\bar\alpha)$ solves the SCWI. In even dimensions, this approach works well, but it is less tractable in odd dimensions, where $\nabla_d^{\frac{d-4}{2}}$ involves fractional powers of a differential operator. In this section, we will show that there exists a local differential operator with the same properties as the one in (\ref{wardSolForm}).
    
    To do this, it is convenient to first understand the solutions of the SCWI in Mellin space, which will be used to build the position space solutions. In~\cite{Zhou:2017zaw}, an approach was developed to rewrite the SCWI in Mellin space as a set of difference equations. The differential operator $z\partial_z$ can be rewritten as:
    \begin{equation}
        z\partial_z = U\partial_U + V\partial_V - \frac{1}{1-z}V\partial_V\;.
    \end{equation}
    Then, we act on $\mathcal{G}$, but leave these derivatives as operators. We now consider the $\alpha\partial_\alpha$ piece. The $\alpha$ derivatives and evaluation at $\alpha=\frac{1}{z}$ generate rational functions of $z$. Since $\mathcal{G}$ is of degree $L$ in $\sigma,\tau$, it follows that we can multiply the result of this evaluation by $z^L(1-z)$ to find a new polynomial in $z$, of degree ${L+1}$. Noting that $U$ and $V$ are invariant under $z\leftrightarrow \bar z$, we can get a second copy of this relation, with $\bar z$ instead of $z$. Adding these two identities together, we find an identity which depends on $z,\bar z$ through functions of the form $z^n+\bar z^n$, which can easily be rewritten in terms of finite linear combinations of terms of the form $U^mV^n$. Then, there exists a simple dictionary between the Mellin and position spaces, where $U\partial_U$, $V\partial_V$ and multiplication by $U^mV^n$ become the operators~\cite{Zhou:2017zaw}:
    \begin{equation}
        \begin{gathered}
            U\partial_U\to \frac{s}{2}\times\;,\quad
            V\partial_V\to \left(\frac{t}{2}-d+2\right)\times\;,\\
            U^mV^n\to\left(d-2-\frac{s}{2}\right)^2_m\left(d-2-\frac{t}{2}\right)^2_n\left(d-2-\frac{u}{2}\right)^2_{-m-n} P_{-2m,-2n}\;,
        \end{gathered}\label{eq:pos_to_mellin}
    \end{equation}
    where $(a)_n$ is the Pochhammer symbol and we defined the shift operator $P_{a,b}$ by:
    \begin{equation}
        P_{a,b}f(s,t) = f(s+a,t+b)\;.
    \end{equation}
    Thus, the superconformal Ward identities imply a set of three difference equations (one for each power of $\bar\alpha$ in a quadratic polynomial), which mix together the coefficients of $\sigma$ and $\tau$ in $\mathcal{M}(s,t;\sigma,\tau)$. In~\cite{Virally:2025nnl}, it was shown that there exists an efficient method to find solutions of the SCWI in Mellin space directly, which, relevant for our purposes, gives an answer which takes the same form in all dimensions. Let us briefly review these solutions. First, we assume that there exists a solution of the SCWI which mirrors the known one in position space, of the form:
    \begin{equation}
        \mathcal{M}(s,t;\sigma,\tau) = \mathcal{D}_d\widetilde{\mathcal{M}}(s,t)\;,\label{eq:full_to_red_max}
    \end{equation}
    where $\mathcal{D}_d$ is now a finite-difference operator with polynomial coefficients, i.e.:
    \begin{equation}
        \mathcal{D}_d = \sum_{a,b,n,m}\sigma^n\tau^m Q_{a,b}^{n,m}(s,t) P_{a,b}\;,
    \end{equation}
    where $Q_{a,b}^{n,m}(s,t)$ are polynomials. The known solutions of the superconformal Ward identity in position space, in even dimensions, can be translated to this form~\cite{Rastelli:2016nze,Rastelli:2017udc,Rastelli:2017ymc}.%
    \footnote{In order to perform this translation, we must know the precise form of the Mellin transform relating the reduced correlator to $\widetilde{\mathcal{M}}(s,t)$. We postpone discussion of this point to later, when we find the solution in position space from the Mellin space operator $\mathcal{D}_d$.}
    Under the assumption that such an operator exists, there must exist a solution of the form:
    \begin{equation}
        \mathcal{M}_{X,Y}(s,t;\sigma,\tau) = \mathcal{D}_dX^sY^t\;,
    \end{equation}
    for any choice of $X,Y$. Shift operators are diagonal in this basis, in the sense that:
    \begin{equation}
        P_{a,b}X^sY^t = (X^aY^b)(X^sY^t)\;.
    \end{equation}
    Thus, we search for a solution of the SCWI of the form:
    \begin{equation}
        \mathcal{M}_{X,Y}(s,t;\sigma,\tau) = \mathcal{Q}_{X,Y}(s,t;\sigma,\tau)X^sY^t\;,
    \end{equation}
    where $\mathcal{Q}_{X,Y}(s,t;\sigma,\tau)$ is a polynomial in $s,t,\sigma,\tau$. Taking the lowest possible order of polynomial solution to this equation, we can immediately read off the shift operators in $\mathcal{D}_d$ from the $X,Y$ dependence, as well as the $Q_{a,b}^{n,m}(s,t)$, such that $\mathcal{D}_d$ is fixed by this procedure, up to redefinitions $\widetilde{\mathcal{M}}(s,t)\to h(s,t)\widetilde{\mathcal{M}}(s,t)$, for which we follow the choice of~\cite{Virally:2025nnl}. It is then simple to check that, for any function $\widetilde{\mathcal{M}}(s,t)$, $\mathcal{M}(s,t;\sigma,\tau)$ as defined by (\ref{eq:full_to_red_max}) solves the SCWI. The explicit operator $\mathcal{D}_d$ for $\langle 2222\rangle$ Mellin amplitudes in all maximally supersymmetric SCFTs can be found in~\cite{Virally:2025nnl}.

    For the reduced Mellin amplitude, the crossing equation (\ref{eq:mellin_crossing_max}) is satisfied if:
    \begin{equation}
        \widetilde{\mathcal{M}}(s,t) = \widetilde{\mathcal{M}}(s,\tilde u) = \widetilde{\mathcal{M}}(\tilde u,t) = \widetilde{\mathcal{M}}(t,s)\;,\label{eq:mellin_crossing_red_max}
    \end{equation}
    where $\tilde u$ is a shifted $u$ variable. In 3d $\mathcal{N}=8$, $s+t+\tilde u=-4$. Because $\mathcal{D}_3$ has a non-trivial kernel, (\ref{eq:mellin_crossing_red_max}) is sufficient but not necessary for crossing symmetry: there always exists a choice of $\widetilde{\mathcal{M}}$ with this property, and any other valid choice differs from it by an element of the kernel. Now, we want to return to position space, and find the analogue of $\mathcal{D}_d$. We define the same relationship as in even dimensions:
    \begin{equation}
        \mathcal{G}(U,V;\sigma,\tau) = \Upsilon_d\left[\widetilde{\mathcal{G}}(U,V)\right]\;,
    \end{equation}
    where we denote the operator by $\Upsilon_d$ instead of writing it in terms of $\nabla_d^{\frac{d-4}{2}}$, since it is now a local differential operator. In order to find $\Upsilon_d$, we first review how we would obtain $\mathcal{D}_d$ from a known $\Upsilon_d$, as done in even dimensions. The first step is to define a Mellin transform which relates $\widetilde{\mathcal{G}}(U,V)$ to $\widetilde{\mathcal{M}}(s,t)$. Then, we act on this transform with $\Upsilon_d$, and use both gamma function identities and shifts of the integration contours to rewrite the integration kernel in the form (\ref{mellinDef}). Then, we can write the transformation as:
    \begin{equation}
        \begin{aligned}
            \Upsilon_d\left[\widetilde{\mathcal{G}}(U,V)\right] ={}& \int \frac{\text{d}s\, \text{d}t}{(4\pi i)^2} U^{\frac{s}{2}}V^{\frac{t}{2}-d+2}\Gamma^2\left(d-2-\frac{s}{2}\right)\Gamma^2\left(d-2-\frac{t}{2}\right)\\
            &\times\Gamma^2\left(d-2-\frac{u}{2}\right)\mathcal{D}_d\widetilde{\mathcal{M}}(s,t)\;,
        \end{aligned}
    \end{equation}
    and from this, we can read off $\mathcal{D}_d$. In order to obtain $\Upsilon_d$, we use the same process. For a choice of the reduced correlator's Mellin transform, we search for a local differential operator such that the right-hand side above (with the known $\mathcal{D}_d$) can be rewritten in this form using gamma function identities and shifts of the integration contour. The existence of such an operator relies on making the correct choice for the reduced correlator's Mellin transform. Our objective is thus to find a Mellin transform which allows us to do this without any non-local operators, and which allows us to write the reduced correlator's block expansion as an expansion in ordinary bosonic blocks. For the 3d $\mathcal{N}=8$ theory, we find that the right Mellin transform is
    \begin{equation}
        \begin{aligned}
            \widetilde{\mathcal{G}}(U,V) ={}& \int \frac{\text{d}s\,\text{d}t}{(4\pi i)^2} U^{\frac{s}{2}+2}V^{\frac{t}{2}+1}\Gamma\left(1-\frac{s}{2}\right)\Gamma\left(-1-\frac{s}{2}\right)\Gamma\left(1-\frac{t}{2}\right)\Gamma\left(-1-\frac{t}{2}\right)\\
            &\times \Gamma\left(1-\frac{\tilde u}{2}\right)\Gamma\left(-1-\frac{\tilde u}{2}\right)\widetilde{\mathcal{M}}(s,t)\;.\label{eq:3d_max_reduced_mellin}
        \end{aligned}
    \end{equation}
    Then, $\Upsilon_3$ can be found, and takes the form of a fourth-order local differential operator in $U,V$:
    \begin{equation}
        \Upsilon_{3} = \sum_{0\leq m+n\leq 4}\frac{p_{m,n}(U,V;\sigma,\tau)}{U^2}\partial_U^m\partial_V^n\;,
    \end{equation}
    where $p_{m,n}(U,V;\sigma,\tau)$ are polynomials. This operator is somewhat complex, so we present it in the attached Mathematica notebook. The operator has a kernel:
    \begin{equation}
        \Upsilon_{3}\left[c_1U+c_2U^2+c_3UV\right]=0\;.
    \end{equation}
    In reduced form, superconformal blocks are given by single bosonic blocks with shifted external dimensions. In this theory, the external dimensions of the blocks are such that $\Delta_{12}=4,\Delta_{34}=0$. We find that the kernel is spanned by the following conformal blocks\footnote{Our convention for conformal blocks is such that $G_{\Delta,\ell}^{\Delta_{12}\Delta_{34}}(z,\bar{z})\sim z^{\frac{\Delta-\ell}{2}}\bar{z}^{\frac{\Delta+\ell}{2}}$ for $0<z\ll\bar z\ll1$; more precisely, $G_{\Delta,\ell}^{\Delta_{12}\Delta_{34}}(z,\bar z)=z^{\frac{\Delta-\ell}{2}}\,k_{\Delta+\ell}(\bar z)+O\big(z^{\frac{\Delta-\ell}{2}+1}\big)$ as $z\to0$, with $k_\beta(\bar z)=\bar z^{\frac{\beta}{2}}\,{}_2F_1\big(\frac{\beta-\Delta_{12}}{2},\frac{\beta+\Delta_{34}}{2};\beta;\bar z\big)$. Blocks of non-integer or negative spin are defined by the same leading term, with the higher orders in $z$ fixed by the Casimir equation, and $\partial_\ell$ acts at this fixed normalization.}:
    \begin{equation}
        \begin{gathered}
        G_{4,0}^{4,0}(U,V) = U^2\;,\quad G_{2,0}^{4,0}(U,V) = \frac{U(3-2U+3V)}{6}\;,\quad
        G_{3,1}^{4,0}(U,V) = U(1-V)\;.
        \end{gathered}
    \end{equation}
    In the reduced correlator, the position space crossing equation (\ref{eq:crossing_pos_max}) becomes:
    \begin{equation}
        \widetilde{\mathcal{G}}(U,V) = V^2\widetilde{\mathcal{G}}\left(\frac{U}{V},\frac{1}{V}\right) = U^3\widetilde{\mathcal{G}}\left(\frac{1}{U},\frac{V}{U}\right) = \frac{U}{V}\widetilde{\mathcal{G}}\left(V,U\right)\pmod{\ker\Upsilon_{3}}\;.
    \end{equation}
    We will show how to deal with the kernel ambiguity at the end of this section. As written, this equation contains logarithmic divergences, due to the twisted sector. Indeed, for any smooth function of $z,\bar z$, we can show from the explicit form of the operator:
    \begin{equation}
        \Upsilon_{3}\left[f_{\text{smooth}}(z,\bar z)\right]\Big|_{z=\bar z=\frac{1}{\alpha}}=0\;.
    \end{equation}
    However, this is not the case when $f$ possesses a logarithmic divergence $\log((z-\bar z)^2)$, in which case it can contribute to the twisted sector. For applications to the numerical bootstrap, it is desirable that the reduced correlator should be smooth when $z\to\bar z$, as we ultimately want to evaluate it and its derivatives at the crossing-symmetric point $z=\bar z = \frac{1}{2}$. Thus, it makes sense to specialize to $1\leftrightarrow3$ crossing and write the full correlator such that the crossing equation is smooth. Under this definition, we have the following:
    \begin{align}
            &\mathcal{G}(U,V;\sigma,\tau) ={}\nonumber\\
            &\quad -\frac{16\lambda^{2}_{1}+\lambda^2_{(B,2)}}{48}\,U\big(\sigma^2(U-1-2V)-3(\tau-1)^2+2\sigma(U-1+V(\tau-1)+\tau(2-U))\big)\nonumber\\
            &\quad-\frac{\lambda^2_{\text{Stress}}}{12}\,U\big(\sigma^2(2U-2-V)+3(\tau-1)^2+2\sigma(V-2+U(\tau-1)+\tau(1-V))\big)\nonumber\\
            &\quad+\frac{\lambda^2_{(B,+)}}{48}\,U\big(\sigma^2(10+8U-7V)+3(\tau-1)^2+2\sigma(V-8+U(\tau-1)+\tau(7-V))\big)\nonumber\\
            &\quad+\Upsilon_3\left[\widetilde{\mathcal{G}}(U,V)\right]\;.
    \end{align}
    
    The reduced block expansion can then be written in the form%
    \footnote{The negative-spin block $G_{3,-2}^{4,0}(U,V)$ can formally be defined by its expansion near $z=\bar z=0$ in lightcone coordinates~\cite{Li:2019cwm}.}:
        \begin{equation}
        \begin{aligned}
            \widetilde{\mathcal{G}}(U,V) ={}& \lambda^2_1\frac{U^2}{48}\log U\\
            &+\lambda^2_{\text{Stress}}\left(\frac{1}{72}G_{3,-2}^{4,0}(U,V) + \frac{1-V}{384}U\log((z-\bar z)^2)\right)\\
            &+\lambda^2_{(B,+)}\left(-\frac{1}{128}\partial_\ell G_{2+\ell,\ell}^{4,0}(U,V)\Bigg|_{\ell=0} + \frac{{U}(3+3V-2U)}{1536}\log((z-\bar z)^2)\right)\\
            &+\lambda^2_{(B,2)}\left(-\frac{1}{768}\partial_\ell G_{4+\ell,\ell}^{4,0}(U,V)\Bigg|_{\ell=0} + \frac{U^2}{1536}\log((z-\bar z)^2)\right)\\
            &+\sum_{\ell\text{ even}} \frac{2\,{n_{\ell+4,\ell+2}}\,\lambda^2_{(A,+)_\ell}}{3(\ell+2)(\ell+1)}\,G_{\ell+4,\ell+2}^{4,0}(U,V)\\
            &-\sum_{\ell\text{ odd}} \frac{4\,{n_{\ell+5,\ell+1}}\,\lambda^2_{(A,2)_{\ell}}}{(\ell+1)(2\ell+5)}\,G_{\ell+5,\ell+1}^{4,0}(U,V)\\
            &+\sum_{\Delta}\sum_{\ell\text{ even}}\frac{{n_{\Delta,\ell}}\,\lambda^2_{\Delta,\ell}}{4(\Delta-\ell)(\Delta-\ell+2)(\Delta+\ell+1)(\Delta+\ell+3)}\,G_{\Delta+4,\ell}^{4,0}(U,V)\;.\label{eq:3d_max_reduced_blocks}
        \end{aligned}
    \end{equation}
    {The factors $n_{\Delta,\ell}=\frac{(1/2)_\ell}{4^{\Delta}\,\ell!}$ convert the unit normalization of our blocks to that of~\cite{Chester:2014fya}.} 
    Here, we subtracted the divergences of each block as $z\to\bar z$ from the blocks themselves, leaving them smooth in the $z\to\bar z$ limit for $0<z,\bar z<1$. The subtractions were chosen such that the $1\leftrightarrow3$ crossing equation for the reduced correlator is left unchanged (the other two crossing relations are not preserved, since $\log((z-\bar z)^2)$ is invariant only under $1\leftrightarrow3$), but we still have the twisted crossing equation (\ref{eq:crossing_twisted_max}). Thus, the reduced crossing relation is given by:
    \begin{equation}
        \begin{gathered}
            4\lambda^2_{\text{Stress}}-5\lambda^2_{(B,+)}+\lambda_{(B,2)}^2+16\lambda_1^2=0\;,\\
            \widetilde{\mathcal{G}}(U,V) - \frac{U}{V}\widetilde{\mathcal{G}}(V,U){}=a\,U(U-V)\;.
        \end{gathered}
    \end{equation}
    The right-hand side of this equation is in the kernel of $\Upsilon_3$. This particular element of the kernel was chosen because it is the most general one which is odd under $1\leftrightarrow 3$ crossing, and the difference of correlators on the left-hand side automatically kills any piece which is even under this transformation. Crossing demands that $a$ should be independent of $U,V$. Thus, we can compute $a$ by evaluating the above on any part of the domain which does not annihilate the right-hand side. A particularly useful choice for numerical computations will be the line $z=\bar z$ near the crossing-symmetric point $z=\bar z=\frac{1}{2}$; at that point itself both sides vanish, and $a$ is fixed by their first derivatives. We find it convenient to define $g(x)$ as the reduced correlator on $z=\bar z=x$:
    \begin{equation}
        g(x) = \widetilde{\mathcal{G}}(x^2,(1-x)^2)\;.
    \end{equation}
    The crossing equation then gives us:
    \begin{equation}
        a = \frac{1}{x^2(2x-1)}g(x)-\frac{1}{(1-x)^2(2x-1)}g(1-x) = 4g'\left(\frac{1}{2}\right)-16g\left(\frac{1}{2}\right)\;,
    \end{equation}
    where the last equality follows from the fact that $a$ is independent of $x$ and from L'H\^opital's rule.

    \section[Flavor correlators with half-maximal supersymmetry]{Flavor correlators with half-maximal\\ supersymmetry}\label{sec:corrHalf}
	\subsection{Full correlator}
	\label{sec:fullHalf}
    We now consider the four-point functions of the scalar superprimaries of half-BPS multiplets $\cD[p], p=2,3,\dots$ in half-maximally supersymmetric SCFTs. These arise in the bulk as excitations on branes wrapping AdS$_{d+1}\times S^3$, and accordingly are often referred to as \textit{gluons}. The scalar superprimary has scaling dimension $\Delta=\frac{d-2}{2}p$, spins $(\frac{p}{2}-1,\frac{p}{2})$ under $SU(2)_L\times SU(2)_R$, and is in the adjoint of the flavor symmetry $G_F$. We use $SU(2)_R$ R-symmetry indices $I$, and $SU(2)_L$ indices $\bar I$, along with $G_F$ indices $A=1,\ldots,\dim(G_F)$.  We will be interested in correlators $\langle 22pp\rangle$ and $\langle pppp\rangle$, which are needed to obtain the CFT data we will use to compute one-loop gluon amplitudes for the $\langle 2222\rangle$ correlator. As in the maximally supersymmetric case, we contract the R-symmetry and isospin indices with auxiliary polarizations, which are now two-component spinors $t^I$ and $\bar t^{\bar I}$ respectively, with the antisymmetric products $t_i\cdot t_j\equiv\epsilon_{IJ}t_i^It_j^J$ and $\bar t_i\cdot \bar t_j\equiv\epsilon_{\bar I\bar J}\bar t_i^{\bar I}\bar t_j^{\bar J}$:
    \begin{equation}
        \mathcal{O}_p^A(x,t,\bar t)\equiv t^{I_1}\ldots t^{I_p}\bar t^{\bar I_1}\ldots\bar t^{\bar I_{p-2}}\mathcal{O}^A_{I_1,\ldots,I_p,\bar I_1,\ldots \bar I_{p-2}}(x)\;.
    \end{equation}
    Then, for a $\langle qqpp\rangle$ correlator with $q\leq p$, which generalizes both $\langle pppp\rangle$ and $\langle 22pp\rangle$, we define:
    \begin{equation}
        \begin{aligned}
        \langle \mathcal{O}_q^A(x_1,t_1,\bar t_1)&\mathcal{O}_q^B(x_2,t_2,\bar t_2)\mathcal{O}_p^C(x_3,t_3,\bar t_3)\mathcal{O}_p^D(x_4,t_4,\bar t_4)\rangle \\
        &= \frac{(t_1\cdot t_2)^{q}(t_3\cdot t_4)^p(\bar t_1\cdot \bar t_2)^{q-2}(\bar t_3\cdot \bar t_4)^{p-2}}{|x_{12}|^{(d-2) q}|x_{34}|^{(d-2) p}}\mathcal{G}^{ABCD}(U,V;\alpha,\beta)\;,
        \end{aligned}
    \end{equation}
    where $\mathcal{G}(U,V;\alpha,\beta)$ encodes the dynamics, the $U,V$ position cross-ratios are defined as in (\ref{eq:cross_ratios_max}) and the R-symmetry and isospin cross-ratios are given by:
    \begin{equation}
        \alpha = \frac{(t_1\cdot t_3)(t_2\cdot t_4)}{(t_1\cdot t_2)(t_3\cdot t_4)}\;,\quad \beta = \frac{(\bar t_1\cdot \bar t_3)(\bar t_2\cdot \bar t_4)}{(\bar t_1\cdot \bar t_2)(\bar t_3\cdot \bar t_4)}\;.
    \end{equation}
    $\mathcal{G}(U,V;\alpha,\beta)$ is a polynomial of order $q$ in $\alpha$, and $q-2$ in $\beta$. In the half-maximal theories, the SCWI read the same as in the maximal case~(\ref{wardGen}). As in the maximally supersymmetric case, the superconformal Ward identities can be solved by expanding into superconformal blocks:
    \begin{equation}
        \mathcal{G}^{ABCD}(U,V;\alpha,\beta) = \sum_{\chi}(\lambda^2_\chi)^{ABCD} \mathcal{G}_\chi(U,V;\alpha,\beta)\;,
    \end{equation}
    where $\chi$ runs over the superconformal multiplets, and each superconformal block $\mathcal{G}_\chi$ solves the SCWI individually. Besides the half-BPS multiplets $\cD[k]$, these are the short multiplets $\cB[k]_\ell$ and the long multiplets $\cL[k]_{\Delta,\ell}$ of~\cite{Chester:2025jxg}, whose superprimary has $SU(2)_R$ spin $\frac{k}{2}$, spin $\ell$, and dimension $\frac{d-2}{2}(k+2)+\ell$ and $\Delta$ respectively (in~\cite{Baume:2019aid} these are $\mathcal{D}[J_R]$, $\mathcal{B}[\ell,J_R]$ and $\mathcal{L}[\Delta,\ell,J_R]$, with $J_R=\frac{k}{2}$). For each $\chi$ there are several multiplets exchanged, in various flavor irreps appearing in the tensor product adj$^2$. Accordingly, each squared OPE coefficient appearing here can be further expanded into projectors onto these irreducible representations,
    \begin{align}
  (\lambda^2_\chi)^{ABCD} = \sum_{\r\in \text{adj}\otimes \text{adj}}(\lambda^2_\chi)_\r P_\r^{ABCD}\,.
  \label{eq: OPE coeff flavor expansion}
\end{align}
We normalize the projectors as
\begin{align}
  P_\r^{ABCD}=P_\r^{CDAB}\,,\qquad P_\r^{ABFE}P_{\r'}^{EFCD}=\delta_{\r\r'}P_\r^{ABCD}\,,\qquad \sum_\r P_\r^{ABCD}=\delta^{AD}\delta^{BC}\,,
  \label{eq: projector normalization}
\end{align}
where complex-conjugate pairs of irreps (such as the $\mathbf{10}\oplus\overline{\mathbf{10}}$ of $SU(3)$) are combined into a single real projector, while the copies of an irrep in the symmetric and antisymmetric parts of $\text{adj}\otimes\text{adj}$ (such as the two adjoints of $SU(N)$, $N\geq 3$) are kept as separate labels $\r$. Then $P_\mathbf{1}^{ABCD}=\delta^{AB}\delta^{CD}/\dim G_F$ (for simple $G_F$), and $P_\r^{BACD}=\pm P_\r^{ABCD}$ for $\r$ in the symmetric (antisymmetric) part of $\text{adj}\otimes\text{adj}$. With this ordering of indices, generalized free field OPE coefficients are positive in every channel, and products of flavor tensors are diagonal: $(X\circ Y)_\r=X_\r Y_\r$, where $(X\circ Y)^{ABCD}\equiv X^{ABFE}Y^{EFCD}$.
The details of the tensor decomposition of $\text{adj}^2$ and the corresponding projectors $P_\r^{ABCD}$ depend on the details of the flavor group $G_F$. We will be agnostic to such details in almost all of what follows. There are some universal facts, however. The identity and stress tensor multiplets appear only in the singlet channel $\lambda_{\cD[0]}^2,\lambda_{\cB[0]_0}^2\propto P_\mathbf{1}$, while $\cD[2]$ appears only in the adjoint that appears in the antisymmetric part of adj$^2$, explicitly $(\lambda_{\cD[2]}^2)^{ABCD}\propto P_\r^{ABCD}=-\frac{1}{2\h}f^{ABE}f^{ECD}$ for this $\r$, where $f^{ABC}$ are the structure constants of $G_F$, normalized as in Section~\ref{sec:largeN}, and $\h$ is its dual Coxeter number. Furthermore $\cB[0]_\ell$ does not appear for $\ell>0$ in an interacting theory. The remaining multiplets appear only through either the symmetric part ($\cD[4]$, $\cB[2]_\ell$ for $\ell$ odd, $\cL[0]_{\Delta,\ell}$ for $\ell$ even) or the antisymmetric part ($\cB[2]_\ell$ for $\ell$ even, $\cL[0]_{\Delta,\ell}$ for $\ell$ odd) of the tensor product $\text{adj}^2$. 
 We will often suppress the flavor indices on the correlator $\cG$ and accordingly on the squared OPE coefficients $\lambda_\chi^2$.

The superconformal blocks can themselves be written as linear combinations of ordinary bosonic blocks:
    \begin{equation}
        \mathcal{G}_\chi(U,V;\alpha,\beta) = \sum_{i,j,\Delta',\ell'}f^{ij}_{\Delta',\ell'}P_i(\alpha)\overline{P}_j(\beta)G_{\Delta',\ell'}(U,V)\;.
    \end{equation}
    Here $P_i(\alpha)$ and $\overline{P}_j(\beta)$ are polynomials of degree $i$ and $j$, corresponding to the exchange of $SU(2)_R$ spin $i$ and $SU(2)_L$ spin $j$. The coefficients $f^{ij}_{\Delta',\ell'}$ can then once again be found by demanding that this should solve the SCWI; since $SU(2)_L$ commutes with the supercharges, they vanish unless $j$ is the $SU(2)_L$ spin of $\chi$. The precise form of the $P_i(\alpha)$ polynomials, as well as the $f^{ij}_{\Delta',\ell'}$ for each exchanged multiplet, can be found in~\cite{Baume:2019aid}.%
    \footnote{Following~\cite{Chester:2025jxg}, where the conversion from the conventions of~\cite{Baume:2019aid} is given, we normalize the $P_i(\alpha)$ such that the highest power of $\alpha$ has coefficient 1, i.e.\ $P_i(\alpha)$ is $\frac{(i!)^2}{(2i)!}$ times the Legendre polynomial of degree $i$ in $2\alpha-1$. The superblocks used below are normalized such that the superprimary, of $SU(2)_R$ spin $i$, dimension $\Delta$ and spin $\ell$, contributes $(-1)^iP_i(\alpha)G_{\Delta,\ell}(U,V)$.}
        In terms of these blocks, the full $\langle 22pp\rangle$ correlator in the half-maximal theories, which is independent of $\beta$, can then be written\footnote{As we are only interested in kinematics for the time being, we keep all $\cB[0]_\ell$, but recall that only $\cB[0]_0$ is present in any interacting theory.}
    \begin{equation}
        \begin{aligned}
            \mathcal{G}(U,V;\alpha) ={}&
            \lambda_{\cD[0]}^2 + \lambda_{\mathcal{D}[2]}^2\mathcal{G}_{\mathcal{D}[2]}(U,V;\alpha) + \lambda_{\mathcal{D}[4]}^2\mathcal{G}_{\mathcal{D}[4]}(U,V;\alpha)\\
            &+\sum_{\ell=0}^\infty \lambda_{\mathcal{B}[0]_\ell}^2\mathcal{G}_{\mathcal{B}[0]_\ell}(U,V;\alpha)+\sum_{\ell=0}^\infty \lambda_{\mathcal{B}[2]_\ell}^2\mathcal{G}_{\mathcal{B}[2]_\ell}(U,V;\alpha)\\
            &+\sum_{\Delta,\ell}\lambda^2_{\Delta,\ell}\mathcal{G}_{\Delta,\ell}(U,V;\alpha)\;,
        \end{aligned}
    \end{equation}
    where the last sum is over the long multiplets $\cL[0]_{\Delta,\ell}$. For $\langle qqpp\rangle$ with $q>2$, further multiplets, such as $\cD[6]$ and $\cB[4]_\ell$, also appear. Note that, for unit-normalized operators $\mathcal{O}_p^A$, we have simply    \begin{align}
  (\lambda_{\cD[0]}^2)^{ABCD} = \delta^{AB}\delta^{CD}\,,
\end{align}
which in the normalization \eqref{eq: projector normalization} is $(\lambda_{\cD[0]}^2)_\r=\dim(G_F)\,\delta_{\r,\mathbf{1}}$.

     We will only use the crossing equation for the $\langle 2222\rangle$ correlator, which is independent of $\beta$. The crossing equation in that setting reads:
    \begin{equation}
        \begin{aligned}
        \mathcal{G}^{ABCD}(U,V;\alpha)={}&\mathcal{G}^{BACD}\left(\frac{U}{V},\frac{1}{V};1-\alpha\right)=U^{d-2}\alpha^2\mathcal{G}^{ACBD}\left(\frac{1}{U},\frac{V}{U};\frac{1}{\alpha}\right)\\ ={}& \frac{U^{d-2}}{V^{d-2}}(\alpha-1)^2\mathcal{G}^{CBAD}\left(V,U;\frac{\alpha}{\alpha-1}\right)\;.\label{eq:crossing_pos_half}
        \end{aligned}
    \end{equation}
    Similarly, we only need the Mellin transform for the $\langle 22pp\rangle$ correlator, which is given by:
    \begin{equation}
        \begin{aligned}
        \mathcal{G}^{ABCD}&(U,V;\alpha) = \int \frac{\text{d}s\,\text{d}t}{(4\pi i)^2}U^{\frac{s}{2}}V^{\frac{t}{2}-\frac{(p+2)(d-2)}{4}}\Gamma\left(d-2-\frac{s}{2}\right)\Gamma\left(\frac{p(d-2)-s}{2}\right)\\
        &\times\Gamma^2\left(\frac{(2+p)(d-2)-2t}{4}\right)\Gamma^2\left(\frac{(2+p)(d-2)-2u}{4}\right)\mathcal{M}^{ABCD}(s,t;\alpha)\;,
        \end{aligned}\label{eq:mellin_22pp}
    \end{equation}
    where $s+t+u=(d-2)(p+2)$, and the left-hand side is understood as the connected correlator, i.e.\ with the disconnected part subtracted. In Mellin space, the crossing equation is given by:
    \begin{equation}
        \begin{aligned}
            \mathcal{M}^{ABCD}(s,t;\alpha) ={}& \mathcal{M}^{BACD}(s,u;{}1-\alpha)=\alpha^2\mathcal{M}^{ACBD}\left(u,t;\frac{1}{\alpha}\right)\\
            ={}&(\alpha-1)^2\mathcal{M}^{CBAD}\left(t,s;\frac{\alpha}{\alpha-1}\right)\;,
        \end{aligned}
    \end{equation}
    where the final two equalities relate $\langle 22pp \rangle$ to $\langle 2p2p \rangle$ and thus only hold for $p=2$.
    
	\subsection{Reduced correlator}
	\label{sec:redHalf}
    As in the maximally supersymmetric case, there is a simpler way to solve the SCWI, without carrying around the R-symmetry cross-ratio. For this section, we specialize to $\langle 2222\rangle$ correlators, which are the simplest case. In Mellin space, explicit solutions to the SCWI are known, and take the form:
    \begin{equation}
        \mathcal{M}(s,t;\alpha) = \mathcal{D}_{d}\widetilde{\mathcal{M}}(s,t)\;.
    \end{equation}
    The operator $\mathcal{D}_{d}$, in dimensions $d=3,5,6$, was found for $\langle 22pp\rangle$ and $\langle 2p2p\rangle$ in~\cite{Chester:2025jxg}, using the methods of~\cite{Virally:2025nnl}. The crossing equation is satisfied if the reduced Mellin amplitude obeys:
    \begin{equation}
        \widetilde{\mathcal{M}}^{ABCD}(s,t) = \widetilde{\mathcal{M}}^{BACD}(s,\tilde u)=\widetilde{\mathcal{M}}^{ACBD}\left(\tilde u,t\right)=\widetilde{\mathcal{M}}^{CBAD}\left(t,s\right)\;,
    \end{equation}
    where $s+t+\tilde u = 4d-12$. As in the maximally supersymmetric case, we now want to find the analogous operator in position space, which we define by:
    \begin{equation}
        \mathcal{G}(U,V;\alpha) = \Upsilon_{d}\left[\widetilde{\mathcal{G}}(U,V)\right]\;.
    \end{equation}
    Our approach to obtaining the local operator is the same as in Section \ref{sec:redMax}. We find that the Mellin transform which connects $\widetilde{\mathcal{G}}(U,V)$ to $\widetilde{\mathcal{M}}(s,t)$ is:
    \begin{equation}
        \begin{aligned}
        \widetilde{\mathcal{G}}(U,V)={}&\int \frac{\text{d}s\,\text{d}t}{(4\pi i)^2}U^{\frac{s}{2}+1}V^{\frac{t}{2}-d+2}\Gamma\left(d-2-\frac{s}{2}\right)\Gamma\left(d-3-\frac{s}{2}\right)\Gamma\left(d-2-\frac{t}{2}\right)\\
        &\times \Gamma\left(d-3-\frac{t}{2}\right)\Gamma\left(\frac{6-2d+s+t}{2}\right)\Gamma\left(\frac{8-2d+s+t}{2}\right)\widetilde{\mathcal{M}}(s,t)\;.
        \end{aligned}\label{eq:reduced_mellin_half}
    \end{equation}
    In this case, $\Upsilon_{d}$ takes a compact form:
    \begin{equation}
        \Upsilon_d = -\frac{8}{U} q^{3-\frac{d}{2}}\mathcal{L}_d q^{\frac{d}{2}-2}\;,\quad q = (1-z\alpha)(1-\bar z\alpha)\;,
    \end{equation}
    where we defined:
    \begin{equation}
        \begin{aligned}
        \mathcal{L}_d ={}& \Big(U(V\partial_V+1)+V(U\partial_U-d+2)\Big)\left(U\partial_U+V\partial_V+2-\frac{d}{2}\right)\\
        &- (V\partial_V+1)(U\partial_U-d+2)\;.
        \end{aligned}
    \end{equation}
    For $d\neq4$, the kernel of this operator is now one-dimensional, and spanned by:
    \begin{equation}
        \Upsilon_d\left[A\frac{U^{d-2}(U^2+(V-1)^2-2U(V+1))^{\frac{3-d}{2}}}{V}\right]=0\;.
    \end{equation}
    We note that this simplifies somewhat in $z,\bar z$ coordinates, as:
    \begin{equation}
        U^2+(V-1)^2-2U(V+1) = (z-\bar z)^2\;.
    \end{equation}
    Furthermore, the kernel is spanned by a single conformal block with $(\Delta_{12},\Delta_{34})=(-2,0)$, which has negative spin for $d>3$:
    \begin{equation}
        \Upsilon_d\left[G^{-2,0}_{d-1,3-d}(U,V)\right]=0\;.
    \end{equation}
    In reduced form, the crossing equation (\ref{eq:crossing_pos_half}) is, up to a term proportional to the kernel:
    \begin{equation}
        \begin{aligned}
        \widetilde{\mathcal{G}}^{ABCD}\left(U,V\right)={}&\frac{1}{V}\widetilde{\mathcal{G}}^{BACD}\left(\frac{U}{V},\frac{1}{V}\right)= U^{d-2}\widetilde{\mathcal{G}}^{ACBD}\left(\frac{1}{U},\frac{V}{U}\right)\\
        ={}& \left(\frac{U}{V}\right)^{d-1}\widetilde{\mathcal{G}}^{CBAD}(V,U)\;.
        \end{aligned}
    \end{equation}
    In $d=5,6$, we choose the reduced correlator to be smooth on the segment $0<z=\bar z<1$; its block expansion is given in (\ref{eq:6d_reduced_correlator}) and (\ref{eq:5d_reduced_correlator}) below. We then find that kernel elements are only required due to the divergence of the negative-spin blocks under $1\leftrightarrow2$ and $1\leftrightarrow4$ crossing. Explicitly, on the Euclidean section $\bar z=z^*$, where the kernel element $K^{\rm E}_d$ defined below is real and positive, one finds:
    \begin{equation}
        \begin{aligned}
            \widetilde{\mathcal{G}}^{ABCD}(U,V)-\left(\frac{U}{V}\right)^{d-1}\widetilde{\mathcal{G}}^{CBAD}(V,U)&=0\;,\\
            \widetilde{\mathcal{G}}^{ABCD}(U,V)-\frac{1}{V}\,\widetilde{\mathcal{G}}^{BACD}\left(\frac{U}{V},\frac{1}{V}\right)&=\xi_d\,\big(\lambda^2_{\mathcal{D}[2]}\big)^{ABCD}\,K^{\rm E}_d(U,V)\;,\\
            \widetilde{\mathcal{G}}^{ABCD}(U,V)-U^{d-2}\,\widetilde{\mathcal{G}}^{ACBD}\left(\frac{1}{U},\frac{V}{U}\right)&=-\xi_d\,\big(\lambda^2_{\mathcal{D}[2]}\big)^{BCAD}\,K^{\rm E}_d(U,V)\;,\\
            K^{\rm E}_d(U,V)&{}\equiv\frac{U^{d-2}}{V}\big(2U(V+1)-U^2-(V-1)^2\big)^{\frac{3-d}{2}}\;,\\
            \xi_5&{}=2\;,\qquad \xi_6=\frac{\pi}{2}\;.
        \end{aligned}
    \end{equation}
    In 3d we will once again need to extract a twisted sector, which we will do in Section \ref{sec:red3Half}. In 6d, the block expansion of the reduced correlator is given by (suppressing flavor indices as ever):
        \begin{equation}
        \begin{aligned}
            \widetilde{\mathcal{G}}(U,V) ={}& -\frac{1}{24}\lambda^2_{\cD[0]}\frac{U}{V}\\
            &+ \frac{1}{8}\lambda^2_{\mathcal{D}[2]}G^{-2,0}_{5,-1}(U,V)\\
            &- \frac{1}{24}\lambda^2_{\mathcal{D}[4]}G^{-2,0}_{8,0}(U,V)\\
            &+\sum_{\ell=0}^\infty\frac{1}{8(\ell+1)}\lambda_{\mathcal{B}[0]_\ell}^2G^{-2,0}_{6+\ell,\ell}(U,V)\\
            &-\sum_{\ell=0}^\infty\frac{\ell+2}{8(\ell+1)(\ell+4)}\lambda_{\mathcal{B}[2]_\ell}^2G^{-2,0}_{9+\ell,\ell+1}(U,V)\\
            &+\sum_{\Delta}\sum_{\ell=0}^\infty \frac{1}{2(\Delta+\ell-2)(6+\ell-\Delta)}\lambda_{\Delta,\ell}^2G_{\Delta+2,\ell}^{-2,0}(U,V)\;,\label{eq:6d_reduced_correlator}
        \end{aligned}
    \end{equation}
    where the negative-spin block has an explicit form best given in terms of $z,\bar z$~\cite{Chang:2017xmr}\footnote{Our reduced correlator is related to the function $b(u,v)$ of~\cite{Chang:2017xmr}, defined by their (3.5) at $\epsilon=k=2$ with $w^{-1}=\alpha-1$, by $\widetilde{\mathcal{G}}(U,V)=-\frac{U^5}{8V}\,b(U,V)$, and (\ref{eq:6d_reduced_correlator}) is the expansion of $b(u,v)$ in the blocks of their (C.8), in our normalization of the blocks and of the OPE coefficients. Since their $\lambda^2_{\mathcal{D}[2]}$ is half of ours, (\ref{eq:6d_neg_spin_block}) equals $-\frac{U^5}{2V}$ times the function $b^{\mathcal{D}[2]}(u,v)$ given explicitly in~\cite{Chang:2017xmr}.}:
    \begin{equation}
        \begin{aligned}
        G^{-2,0}_{5,-1}(z,\bar z) ={}& \frac{2z \bar z}{(z-1)(\bar z-1)(z-\bar z)^3}\Big(z^3\bar z^3(\log\bar z-\log z)\\
        &+ \bar z^2(z-1)(z^2\bar z+z\bar z+\bar z-3z)\log(1-z)\\
        &-z^2(\bar z-1)(z\bar z^2+z\bar z+z-3\bar z)\log(1-\bar z)-z\bar z(z-\bar z)(z\bar z-z-\bar z)\Big)\;.\label{eq:6d_neg_spin_block}
        \end{aligned}
    \end{equation}
    The leading term and the Casimir equation fix this block only up to a multiple of the kernel block $G^{-2,0}_{5,-3}$; the expression above is the unique choice that is smooth on the segment $0<z=\bar z<1$. In 5d, there is a similar block expansion:
        \begin{equation}
        \begin{aligned}
            \widetilde{\mathcal{G}}(U,V) ={}& -\frac{1}{8}\lambda^2_{\mathcal{D}[0]}\frac{U}{V}\\
            &+ \frac{1}{2}\lambda^2_{\mathcal{D}[2]}\left(G^{-2,0}_{4,-1}(U,V)- \frac{2U^3}{V(U^2+(V-1)^2-2U(V+1))}\right)\\
            &- \frac{1}{8}\lambda^2_{\mathcal{D}[4]}G^{-2,0}_{6,0}(U,V) \\
            &+\sum_{\ell=0}^\infty\frac{1}{4(\ell+1)}\lambda_{\mathcal{B}[0]_\ell}^2G^{-2,0}_{5+\ell,\ell}(U,V)\\
            &-\sum_{\ell=0}^\infty\frac{2\ell+3}{8(\ell+1)(\ell+3)}\lambda_{\mathcal{B}[2]_\ell}^2G^{-2,0}_{7+\ell,\ell+1}(U,V)\\
            &+\sum_{\Delta}\sum_{\ell=0}^\infty \frac{1}{2(\Delta+\ell-1)(4+\ell-\Delta)}\lambda_{\Delta,\ell}^2G_{\Delta+2,\ell}^{-2,0}(U,V)\;,\label{eq:5d_reduced_correlator}
        \end{aligned}
    \end{equation}
    where we added an element of the kernel to the negative-spin block to make the correlator smooth on the segment $0<z=\bar z<1$. This time, we find that the negative-spin block has the following representation in terms of Appell's $F_1$ function:
    \begin{equation}
        G^{-2,0}_{4,-1}(z,\bar z) = -\frac{(z\bar z)^{5/2}}{(1-z)(1-\bar z)(z-\bar z)^2}\left[z\bar zF_1\left(\frac{5}{2};\frac{1}{2},\frac{1}{2};4;z,\bar z\right) -(z+\bar z)F_1\left(\frac{3}{2};\frac{1}{2},\frac{1}{2};3;z,\bar z\right) \right]\;.
    \end{equation}
	
	\subsection{Reduced block expansion for \texorpdfstring{$d=3$}{d=3}}
	\label{sec:red3Half}
    In three dimensions, there is a twisted sector, much like in the case of 3d $\mathcal{N}=8$. This time there is no second R-symmetry cross-ratio $\bar\alpha$, so that after twisting the correlator is constant on each of the intervals $z<0$, $0<z<1$ and $z>1$. When we set $z=\bar z = \frac{1}{\alpha}$, the SCWI give us:
    \begin{equation}
        \frac{\text{d}}{\text{d}z}\left[\mathcal{G}(U,V;\alpha)\Big|_{\bar z=z,\ \alpha=\frac{1}{z}}\right]=\left(\partial_z+\partial_{\bar z}-\alpha^2\partial_\alpha\right)\mathcal{G}(U,V;\alpha)\Big|_{z=\bar z=\frac{1}{\alpha}}=0\;.
    \end{equation}
    In terms of $\Upsilon_{3}$, once more, for any smooth function of $z,\bar z$, the topological twist vanishes:
    \begin{equation}
        \Upsilon_{3}\left[f_{\text{smooth}}(z,\bar z)\right]\Big|_{z=\bar z=\frac{1}{\alpha}}=0\;.
    \end{equation}
    The relevant divergent function to reproduce the topologically twisted sector is any constant multiple of $\frac{U}{V}\log((z-\bar z)^2)$. As in the maximally supersymmetric case, we want to write the 3d conformal block decomposition in a way that is smooth in the $z\to\bar z$ limit, in order to enable evaluation at the crossing-symmetric point. Once more, we thus specialize to $1\leftrightarrow3$ crossing, and write the full correlator as%
    \footnote{Here $\sgn(z)$ is $+1$ for $0<z,\bar z<1$ and $-1$ for $z,\bar z<0$, with $\sqrt{U}>0$ in both regions; at $z=\bar z=\frac{1}{\alpha}$ it is the sign function in (5.3) of~\cite{Chester:2023qwo}.}:
    \begin{equation}
        \mathcal{G}(U,V;\alpha) = \left(\lambda^2_{\mathcal{D}[0]}-\sgn(z)\lambda_{\mathcal{D}[2]}^2+\lambda^2_{\mathcal{D}[4]}\right)U\alpha^2 + \Upsilon_3\left[\widetilde{\mathcal{G}}(U,V)\right]\;,
    \end{equation}
        where the reduced correlator admits the following block decomposition:
    \begin{equation}
        \begin{aligned}
            \widetilde{\mathcal{G}}(U,V) ={}& -\frac{1}{4}\lambda^2_{\mathcal{D}[0]}\frac{U}{V}\log U\\
            &+ \frac{1}{2}\lambda^2_{\mathcal{D}[2]}\left(G^{-2,0}_{2,-1}(U,V) + \sgn(z)\frac{U}{4V}\log((z-\bar z)^2)\right)\\
            &+ \frac{1}{4}\lambda^2_{\mathcal{D}[4]}\left(\partial_\ell G_{2+\ell,\ell}^{-2,0}(U,V)\Bigg|_{\ell=0}-\frac{U}{2V}\log((z-\bar z)^2)\right)\\
            &-\sum_{\ell=0}^\infty\frac{1}{4(\ell+1)}\lambda_{\mathcal{B}[0]_\ell}^2G^{-2,0}_{3+\ell,\ell}(U,V)\\
            &+\sum_{\ell=0}^\infty\frac{2\ell+1}{8(\ell+1)^2}\lambda_{\mathcal{B}[2]_\ell}^2G^{-2,0}_{3+\ell,\ell+1}(U,V)\\
            &+\sum_{\Delta}\sum_{\ell=0}^\infty \frac{1}{2(\Delta+\ell+1)(\ell-\Delta)}\lambda_{\Delta,\ell}^2G_{\Delta+2,\ell}^{-2,0}(U,V)\;.
        \end{aligned}\label{eq:3d_reduced_correlator}
    \end{equation}
    As in the maximally supersymmetric case, the negative-spin block is best defined from the lightcone expansion~\cite{Li:2019cwm}. The function $\frac{U}{V}\log((z-\bar z)^2)$, which carries the twisted sector, is invariant under the $1\leftrightarrow3$ transformation of Section~\ref{sec:redHalf} but not under the other two, so that with the twisted sector split off only the $1\leftrightarrow3$ relation holds up to the kernel. Once again, we have separate crossing equations for the part which contributes to the topological twist, and for the reduced correlator:
    \begin{equation}
        \begin{gathered}
           \left( \lambda^2_{\mathcal{D}[0]}-\lambda_{\mathcal{D}[2]}^2+\lambda^2_{\mathcal{D}[4]}\right)^{ABCD}=\left(\lambda^2_{\mathcal{D}[0]}-\lambda_{\mathcal{D}[2]}^2+\lambda^2_{\mathcal{D}[4]}\right)^{CBAD}\;,\\
            \widetilde{\mathcal{G}}^{ABCD}(U,V) - \frac{U^2}{V^2}\widetilde{\mathcal{G}}^{CBAD}(V,U) = A^{ABCD}\frac{U}{V}\;.
        \end{gathered}
    \end{equation}
    As in the case of 3d $\mathcal{N}=8$, we must have a piece of the kernel which is odd under crossing, which implies that $A^{ABCD}=-A^{CBAD}$. Crossing also requires that $A^{ABCD}$ should be independent of $U,V$. Once more, the most convenient place to evaluate $A^{ABCD}$ will be at $z=\bar z=\frac{1}{2}$, which gives us:
    \begin{equation}
        A^{ABCD} = \widetilde{\mathcal{G}}^{ABCD}\left(\frac{1}{4},\frac{1}{4}\right) -\widetilde{\mathcal{G}}^{CBAD}\left(\frac{1}{4},\frac{1}{4}\right)\;.
    \end{equation}
	
	\section{One-loop gluon scattering}\label{sec:1loop}	
	In this section, we put our reduced position space formalism to use, by computing for the first time the one-loop gluon amplitude in $d=3,5,6$, generalizing the $d=4$ result of \cite{Alday:2021ajh}. We first briefly review the object we would like to compute, before applying the method of \cite{Alday:2021ajh} to compute it in our reduced position space. We will then extract one-loop CFT data of various short and lowest twist long multiplets from this amplitude via Lorentzian inversion, before finally verifying that the flat space limit of our one-loop amplitude matches the expected result.

	\subsection{Large \texorpdfstring{$N$}{N} expansion}
	\label{sec:largeN}

At large $N$, the half-maximal theories in $d=3,4,5,6$ admit a double expansion in terms of the inverse central charge $1/c_T$, which in the bulk controls the graviton self-coupling as well as the coupling between gravitons and gluons, and the inverse flavor central charge $1/c_J$, which in the bulk controls the gluon self-coupling. We normalize each of these central charges such that we have the stress tensor and current two-point functions
\begin{equation}
\begin{aligned}
\label{charges}
	\left< T_{ab}(z) T_{cd}(0)\right> =&\, \frac{\Gamma(d/2)^2}{4\pi^{d}} \frac{c_T}{z^{2d}} \left( I_{a(c}I_{d)b} - \text{trace} \right)\,,\\
	\left< J^A_{a}(z) J^B_{b}(0)\right> =&\, \frac{\Gamma(d/2)^2}{4\pi^{d}}\frac{c_J}{z^{2(d-1)}}  \delta^{AB} I_{ab}\,,	
\end{aligned}
\end{equation}
where $I_{ab}=\delta_{ab}-2z_az_b/z^2$, and we normalize Lie algebra generators such that the norm-squared of the longest root of $G_F$ is $\psi^2=2$. Equivalently, the structure constants satisfy
\begin{align*}
  f^{ACD}f^{BCD}=2h^\vee \delta^{AB}\,,
\end{align*}
 where $h^\vee$ is the dual Coxeter number of $G_F$.
 
 Schematically, if the gauge theory on AdS$_{d+1}\times S^3$ has a single UV scale, then crossing, analyticity and the flat space limit \cite{Chester:2018aca,Chester:2023qwo,Chester:2025jxg} constrain the large $N$ expansion of the reduced correlator to take the form
\begin{align}
  d=3{:}\quad \widetilde{\cG} &= \widetilde{\cG}_\text{GFFT} + \frac{1}{c_J}\, \widetilde{\cG}_F + \frac{1}{c_J^2}\,\widetilde{\cG}_{F|F} + \frac{1}{c_J^{7/3}}\,\widetilde{\cG}_{F^4} + O\big(N^{-3/2}\log N\big)\,, \nn\\
  d=5{:}\quad \widetilde{\cG} &= \widetilde{\cG}_\text{GFFT} + \frac{1}{c_J}\, \widetilde{\cG}_F + \frac{1}{c_T}\big[\widetilde{\cG}_R+\dots\big] + \frac{1}{c_J^{9/5}}\,\widetilde{\cG}_{F^4} + \frac{1}{c_J^2}\big[\widetilde{\cG}_{F|F} + \widetilde{\cG}'_{F^4}\big] + O\big(N^{-33/10}\big)\,, \nn\\
  d=6{:}\quad \widetilde{\cG} &= \widetilde{\cG}_\text{GFFT} + \frac{1}{c_J}\, \widetilde{\cG}_F + \frac{1}{c_T}\big[\widetilde{\cG}_R+\dots\big] + \frac{1}{c_J^{5/3}}\,\widetilde{\cG}_{F^4} + \frac{1}{c_J^2}\big[\widetilde{\cG}_{F|F} + \log c_J\,\widetilde{\cG}_{\log} + \widetilde{\cG}_{D^2F^4}\big] \nn\\
  &\qquad + O\big(N^{-13/3}\big)\,,
\end{align}
where $c_J\sim N^{1/2},N^{3/2},N^2$ and $c_T\sim N^{3/2},N^{5/2},N^3$ in $d=3,5,6$ respectively, with $N$ the number of M2, D4 and M5 branes. Here $\widetilde{\cG}_\text{GFFT}$ is the generalized free field theory (GFFT) term, $\widetilde{\cG}_F$ and $\widetilde{\cG}_{F|F}$ are the tree-level and one-loop gluon terms, while $\widetilde{\cG}_R$ is graviton exchange, computed for $d=3,6$ in \cite{Chester:2025jxg} (see also \cite{Chester:2023qwo}), with the ellipses denoting subleading corrections to it. The terms $\widetilde{\cG}_{F^4}$, $\widetilde{\cG}'_{F^4}$ and $\widetilde{\cG}_{D^2F^4}$ are contact terms, whose reduced Mellin amplitudes are polynomials in $s,t$ of degree zero for the first two and at most one for the third, with coefficients that are not fixed by the bootstrap. Finally, $\widetilde{\cG}_{\log}$ arises from the logarithmically divergent ten-dimensional box diagram. The contact terms and $\widetilde{\cG}_{\log}$ only affect the CFT data of low spins. A similar expansion can be written for the full correlator.

In $d=3$ the contact term $\widetilde{\cG}_{F^4}$ is known to vanish \cite{Chester:2023qwo}, while graviton exchange first appears at order $1/c_T\sim N^{-3/2}$.

Following the 4d case in \cite{Alday:2021ajh}, we can then determine the one-loop contribution $\widetilde{\cG}_{F|F}$ to $\langle 2222\rangle$ given knowledge of the generalized free field contribution $\cG_\text{GFFT}$ to $\langle pppp\rangle$ for every $p=2,3,\dots$, along with the tree-level contribution $\cG^{(d)}_{F,p}$ to $\langle 22pp\rangle$ for every $p=2,3,\dots$.

    \subsection{Data from GFFT and tree-level gluon amplitudes}\label{sec:gfft_tree_data}
First, we consider the GFFT correlators. We need them only for $\langle pppp\rangle$, where they take a simple form in position space:
    \begin{equation}
        \begin{aligned}
            \mathcal{G}^{ABCD}_{\text{GFFT}}(U,V;\alpha,\beta) ={}& \delta^{AB}\delta^{CD} + \delta^{AD}\delta^{BC}(\alpha-1)^{p}(\beta-1)^{p-2}\frac{U^{\frac{d-2}{2}p}}{V^{\frac{d-2}{2}p}}+ \delta^{AC}\delta^{BD}\alpha^p\beta^{p-2}U^{\frac{d-2}{2}p}\;.
        \end{aligned}
    \end{equation}
    From these, we want to obtain the GFFT OPE coefficients. Crossing symmetry relates the different channels, such that we only need to obtain these in one channel. We will choose the $t$ channel, where the OPE coefficients $A_{p,n,\ell}^{(d,0)}$ are defined by:
    \begin{equation}
        (\alpha-1)^p(\beta-1)^{p-2}\frac{U^{\frac{d-2}{2}p}}{V^{\frac{d-2}{2}p}}\bigg|_{SU(2)_L\text{ singlet}} = \sum_{n,\ell}A_{p,n,\ell}^{(d,0)}\,\mathcal{G}_{p(d-2)+2n+\ell,\ell}(U,V;\alpha)+\dots\;,
    \end{equation}
    where the subscript denotes the projection onto the $SU(2)_L$ singlet, under which $(\beta-1)^{p-2}$ is replaced by $\frac{(-1)^p}{p-1}$; $\mathcal{G}_{\Delta,\ell}(U,V;\alpha)$ is the superblock of the long multiplet $\cL[0]_{\Delta,\ell}$, normalized such that the conformal block of its superprimary has unit coefficient; and the dots denote the superblocks of all other multiplets.
Explicitly, the OPE coefficients are:
    \begin{equation}
        \begin{aligned}
            A^{(d,0)}_{p,n,\ell} ={}& 2^{-2+d-2\ell-4n+4p-2dp}\frac{(d-2)^2\pi\Gamma\left(\frac{d}{2}+\ell\right)\Gamma\left(n+\frac{1}{2}(d-2)(p-1)\right)}{\Gamma (\ell+1) \Gamma (n+1) \Gamma\left(\frac{d}{2}+\ell+n\right) \Gamma\left(n+\frac{1}{2} (d-2) (p-1)+\frac{1}{2}\right)}\\
            &\times\frac{\Gamma\left(3+n+d(p-1)-2p\right)\Gamma\left(2+\ell+n+d\left(p-\frac{1}{2}\right)-2p\right)}{\Gamma^2 \left(\frac{1}{2} d (p-1)-p+2\right) \Gamma^2\left(\frac{1}{2} (d-2) p\right) \Gamma \left(\ell+2n+d \left(p-\frac{1}{2}\right)-2 p+2\right)}\\
            &\times\frac{\Gamma \left(\ell+n+\frac{1}{2} (d-2) p\right) \Gamma(\ell+2 n+(d-2) p+1)}{((d-2) p+2 n) (d (p+1)+2 (\ell+n-p-1))\Gamma\left(\ell+n+\frac{1}{2} ((d-2) p+1)\right)}\;,
        \end{aligned}\label{eq:gfft_ope}
    \end{equation}
    and can also be found in the attached Mathematica notebook. We now need to compute anomalous dimensions and OPE coefficients from the tree-level gluon amplitudes. Specifically, when either $d$ or $p$ is even, we obtain the anomalous dimensions, while when both $d$ and $p$ are odd we will instead obtain the tree-level OPE coefficient. The tree-level amplitudes were obtained in~\cite{Alday:2021odx} in Mellin space for all $\langle 22pp\rangle$ correlators, where they take the form:
    \begin{equation}
        \mathcal{M}_{F,p}(s,t;\alpha) = \frac{4(d-1)}{(d-2)}\big(\mathbf{c}_s\mathcal{M}^s_{p}(s,t;\alpha)+\mathbf{c}_t\mathcal{M}_{p}^t(s,t;\alpha)+\mathbf{c}_u\mathcal{M}_{p}^u(s,t;\alpha)\big)\;,
        \label{eq:gluon_tree_mellin}
    \end{equation}
    where $\mathcal{M}_{F,p}$ is the coefficient of $1/c_J$ in the $\langle 22pp\rangle$ Mellin amplitude, so the prefactor is $c_J$ times the squared OPE coefficient of the flavor current multiplet, $\frac{4(d-1)}{(d-2)\,c_J}$, which is fixed by the flavor Ward identity~\cite{Chang:2017xmr,Alday:2021odx}, and $\mathbf{c}_a$ are flavor structures,
        \begin{equation}
        \mathbf{c}_s^{ABCD} = f^{ABE}f^{ECD}\;,\quad \mathbf{c}_t^{ABCD} = f^{ADE}f^{EBC}\;,\quad \mathbf{c}_u^{ABCD} = f^{ACE}f^{EDB}\;,
    \end{equation}
    which obey the Jacobi identity:
    \begin{equation}
        \mathbf{c}_s+\mathbf{c}_t+\mathbf{c}_u=0\;.
    \end{equation}
    In the conventions of Section~\ref{sec:fullHalf}, the squared OPE coefficient of the flavor current multiplet in $\langle 2222\rangle$ is $(\lambda^2_{\cD[2]})^{ABCD}=-\frac{4(d-1)}{(d-2)\,c_J}\,\mathbf{c}_s^{ABCD}$, which is $\frac{8(d-1)h^\vee}{(d-2)\,c_J}$ times the projector onto the adjoint in the antisymmetric part of $\text{adj}\otimes\text{adj}$. The $\mathcal{M}_{p}^a(s,t;\alpha)$ are the $\langle 22pp\rangle$ amplitudes $\mathcal{M}_{s,t,u}$ of~\cite{Alday:2021odx}, given in equations (5.5), (4.16) and (4.21) of that reference for $d=3,5,6$, divided by their overall coupling; this coupling, given in equation (4.4) of~\cite{Alday:2021odx}, is $\frac{4(d-1)}{(d-2)\,c_J}$, i.e.\ $\frac{1}{c_J}$ times the prefactor in (\ref{eq:gluon_tree_mellin}). We are now interested in computing CFT data from these. The first step will be to return to position space. We do not need to develop a reduced correlator for $\langle 22pp\rangle$ to extract the CFT data. Instead, we note that, in the $\langle 22pp\rangle$ correlator, the $\alpha^2$ coefficient of any superconformal block contains at most a single bosonic block. In the normalization of~\cite{Chester:2025jxg}:
        \begin{equation}
        \mathcal{G}_{\Delta,\ell}(U,V;\alpha)\Bigg|_{\alpha^2} = \frac{(\Delta+\ell)(\ell-\Delta+d-2)}{(\ell+2d-\Delta-6)(\Delta+\ell+4-d)}G_{\Delta+2,\ell}(U,V)\;.
    \end{equation}
    We can use (\ref{eq:mellin_22pp}) to transform the Mellin amplitudes in (\ref{eq:gluon_tree_mellin}) to position space, and then expand in conformal blocks to obtain the CFT data. We give some details of this transformation in Appendix \ref{sec:tree_position}. We then use the Jacobi identity to set $\mathbf{c}_u=-\mathbf{c}_s-\mathbf{c}_t$. By crossing symmetry, it is sufficient to consider the coefficient of $\mathbf{c}_s$. When either $d$ or $p$ is even, we only need the coefficient of $\log U$, which gives us the anomalous dimensions weighted by OPE coefficients, $\omega^{(d)}_{p,n,\ell}$, defined by:
        \begin{equation}
        \begin{aligned}
            \mathcal{G}_{F,p}^{(d)}(U,V;\alpha)\Big|_{\alpha^2\mathbf{c}_s\log U} ={}& \sum_{n,\ell} \frac{(d (p-1)+2 (n-p+1)) (dp+2 \ell+2 n-2 p)}{(d(p-2)+2 (n-p+3)) (d (p-1)+2(\ell+n-p+2))}\\
            &\times\omega^{(d)}_{p,n,\ell}\,G_{p(d-2)+2n+\ell+2,\ell}(U,V)\;,
        \end{aligned}
    \end{equation}
    where the prefactor is the $\alpha^2$ coefficient above, evaluated at the superprimary dimension $\Delta=p(d-2)+2n+\ell$.
When both $d$ and $p$ are odd, the double traces $[\cO_p\cO_p]_{n,\ell}$ have odd twist and are absent from the generalized free $\cO_2\times\cO_2$ OPE, so there is no $\log U$ term at this order.
 Instead, we compute the terms with half-integer powers of $U$ in the $\alpha^2\mathbf{c}_s$ component, denoted by $U^{\mathbb{Z}+\frac12}$, and extract the tree-level OPE coefficients $A^{(d,1)}_{p,n,\ell}$, defined as:
        \begin{equation}
        \begin{aligned}
            &\mathcal{G}_{F,p}^{(d)}(U,V;\alpha)\Big|_{\alpha^2\mathbf{c}_s,\,U^{\mathbb{Z}+\frac12}} =\sum_{n,\ell} \frac{(d (p-1)+2 (n-p+1)) (dp+2 \ell+2 n-2 p)}{(d(p-2)+2 (n-p+3)) (d (p-1)+2(\ell+n-p+2))}\\
            &\hspace{70mm}\times A^{(d,1)}_{p,n,\ell}\,G_{p(d-2)+2n+\ell+2,\ell}(U,V)\;.
        \end{aligned}
    \end{equation}
    In general, we find that the tree-level OPE coefficients are related to the analytic continuation of $\omega^{(d)}_{p,n,\ell}$ in $p$ by $A^{(d,1)}_{p,n,\ell} = -i\pi\omega^{(d)}_{p,n,\ell}$. The $\omega^{(d)}_{p,n,\ell}$ take the explicit form below, where each ${}_4F_3$ denotes the sum of the first $n+1$ terms of its hypergeometric series (for $p=2$ in $d=3,5$ a lower parameter coincides with the upper parameter $-n$):
    \begin{equation}
        \begin{aligned}
            \omega^{(3)}_{p,n,\ell} ={}& (-1)^\ell 2^{3-2\ell-4n-p}e^{\frac{ip\pi}{2}}\frac{\Gamma \left(\ell+\frac{3}{2}\right) \Gamma\left(n+\frac{1}{2}\right) \Gamma (2 n+p-1)}{\pi^2\Gamma(n+1)\Gamma\left(\frac{3}{2}+\ell+n\right)\Gamma\left(\frac{1+p}{2}+\ell+n\right)}\\
            &\times\frac{\Gamma\left(\ell+n+\frac{p}{2}\right) \Gamma\left(\ell+n+p+\frac{1}{2}\right)}{\Gamma(p-1)\Gamma(\frac{1}{2}+\ell+2n+p)}\\
            &\times\,_4F_3\biggl(1+\ell,-n,-\ell-2n-p,\frac{1-p}{2}-n;\\
            &\hspace{18mm}\frac{1}{2}-n,1-2n-p,1-n-\frac{p}{2};1\biggr)\;,\\
            \omega^{(5)}_{p,n,\ell} ={}& (-1)^{\ell+n}3^{-\frac{3 p}{2}} 2^{-2 (\ell+3 n+3 p-4)} e^{\frac{3 i \pi  p}{2}}\frac{\Gamma \left(\ell+\frac{5}{2}\right) (2 n+3 p-2)^2 (2 \ell+2n+3 p+1)}{\Gamma (n+1) \Gamma\left(\frac{p}{2}-\frac{1}{6}\right) \Gamma\left(\frac{p}{2}+\frac{1}{6}\right) \Gamma\left(\ell+n+\frac{5}{2}\right)}\\
            &\times\frac{\Gamma (n+3 p-2)\Gamma (2 n+3 p-3) \Gamma \left(\ell+n+\frac{3p}{2}\right)\Gamma\left(\ell+n+3p-\frac{1}{2}\right)}{\Gamma \left(\frac{3 p}{2}\right) \Gamma\left(\frac{p+1}{2}\right) \Gamma (3 p-4) \Gamma\left(\ell+n+\frac{3 p}{2}-\frac{1}{2}\right) \Gamma\left(\ell+2 n+3 p-\frac{1}{2}\right)}\\
            &\times\, _4F_3\biggl(-n,-n-\frac{3p}{2}+\frac{3}{2},-\ell-n-\frac{3 p}{2},\ell+n+\frac{3p}{2}+1;\\
            &\hspace{18mm}{-}n-\frac{3 p}{2}+1,-n-\frac{3p}{2}+3,\frac{3 p}{2}-\frac{1}{2};1\biggr)\;,\\
            \omega^{(6)}_{p,n,\ell} = {}&(-1)^{\ell+1}5\cdot 2^{4-2\ell-4n-8p}\frac{\pi  (n+2 p-3) (\ell+n+2 p-1) \Gamma (n+2 p) \Gamma (n+4p-3)}{(\ell+3) \Gamma (n+1) \Gamma (2 p) \Gamma (2 p-2) (\ell+2n+4 p-2)}\\
            &\times\frac{\Gamma (\ell+n+2 p+2)}{\Gamma \left(n+\frac{1}{2} (4 p-3)\right) \Gamma\left(n+\frac{1}{2} (4 p+1)+\ell \right)}\Bigl(2 n^2+n (4 p-2) \ell +n (16 p-10)\\
            &\hspace{18mm}+\left(8 p^2-2p-1\right) \ell +24 p^2+(2 p-1) \ell ^2-24 p+6\Bigr)\;.
        \end{aligned}
    \end{equation}

		\subsection{Mellin amplitude}
	\label{sec:AnswerMel}
        We now have all of the data we need to compute the one-loop amplitudes for gluon scattering in AdS. We want to compute the coefficient of $\log^2U$ for the reduced amplitude in position space, and then obtain the corresponding Mellin amplitude. Crossing in Mellin space then provides us with the complete reduced amplitude. We firstly have
    \begin{equation}\label{eq:blah}
        \widetilde{\mathcal{G}}_{F|F}(U,V)\Big|_{\log^2U} = \Big[(\mathbf{d}_{st}+\mathbf{d}_{su})\,\widetilde{\mathcal{F}}_{+}(U,V)+(\mathbf{d}_{st}-\mathbf{d}_{su})\,\widetilde{\mathcal{F}}_{-}(U,V)\Big]\;.
    \end{equation}
    The flavor structures are:
    \begin{align}
        \d_{st}^{ABCD} &= f^{EAF}f^{FBG}f^{GCH}f^{HDE},\nn\\
         \d_{tu}^{ABCD} &= f^{EAF}f^{FCG}f^{GBH}f^{HDE},			\nn\\
 \d_{su}^{ABCD} &= f^{EAF}f^{FBG}f^{GDH}f^{HCE}.
    \end{align}
In \eqref{eq:blah}, we defined the following functions, which separate out even and odd spins:
    \begin{equation}
        \widetilde{\mathcal{F}}_{\pm}(U,V) = -\sum_{n=0}^\infty\sum_{\ell=0}^\infty \left(\frac{1\pm (-1)^\ell}{2}\right)\frac{F_{n,\ell}}{4(n+1)(d+2n+2\ell)}\,G_{2(d-2)+2n+\ell+2,\ell}^{-2,0}(U,V)\;.
    \end{equation}
The double traces contributing to $F_{n,\ell}$ are degenerate at large $c_J$. As in 4d~\cite{Alday:2021ajh}, the average of their squared anomalous dimensions can be computed separately in each flavor irrep and, by large $c_J$ factorization, needs only the GFFT data of $\langle pppp\rangle$ and the tree-level data of $\langle 22pp\rangle$. We find
\begin{equation}
        F_{n,\ell} = \sum_{\substack{p=2\\ p\ \text{even if $d$ is odd}}}^{2+\frac{2n}{d-2}}\frac{\Big(\omega^{(d)}_{p,\,n-\frac{(d-2)(p-2)}{2},\,\ell}\Big)^2}{2\,A^{(d,0)}_{p,\,n-\frac{(d-2)(p-2)}{2},\,\ell}}\;.
    \end{equation}
    The odd dimensions have an additional contribution: $\widetilde{\mathcal{G}}_{F|F}(U,V)$ also contains terms with half-integer powers of $U$ and no $\log U$, which are given by the right-hand side of \eqref{eq:blah} with $\widetilde{\mathcal{F}}_{\pm}(U,V)$ replaced by $\widetilde{\mathcal{F}}_{\pm,\text{odd}}(U,V)$, where $\widetilde{\mathcal{F}}_{\pm,\text{odd}}(U,V)$ comes from exchanges of operators of odd twist.
These are the double traces $[\cO_p\cO_p]_{n,\ell}$ with odd $p$, which are absent from the GFFT $\cO_2\times\cO_2$ OPE. Their squared OPE coefficients in $\langle 2222\rangle$ therefore first appear at order $1/c_J^2$ and are fixed by the tree-level coefficients $A^{(d,1)}_{p,n,\ell}$ in $\langle 22pp\rangle$, and their contribution has a non-zero double discontinuity after crossing.
This is the flavored extension, also to $d=5$, of the 3d $\mathcal{N}=8$ graviton calculation in~\cite{Alday:2022rly} (see eq.~(2.23) there). We find the following expression:
    \begin{equation}
        \widetilde{\mathcal{F}}_{\pm,\text{odd}}(U,V) = -\sum_{n=0}^\infty\sum_{\ell=0}^\infty \frac{\big(1\pm (-1)^\ell\big)\,F_{n,\ell,\text{odd}}}{8(d+2n)(d+n+\ell-1)}\,G_{3(d-2)+2n+\ell+2,\ell}^{-2,0}(U,V)\;,
    \end{equation}
where the $F_{n,\ell,\text{odd}}$ have a similar expression, but now involve the tree-level OPE coefficients $A^{(d,1)}_{p,n,\ell}$ instead of the $\omega^{(d)}_{p,n,\ell}$:
\begin{equation}
        F_{n,\ell,\text{odd}} = \sum_{\substack{p=3\\ p\ \text{odd}}}^{3+\frac{2n}{d-2}}\frac{\Big(A^{(d,1)}_{p,\,n-\frac{(d-2)(p-3)}{2},\,\ell}\Big)^2}{A^{(d,0)}_{p,\,n-\frac{(d-2)(p-3)}{2},\,\ell}}\;.
    \end{equation}
    Using (\ref{eq:reduced_mellin_half}), we can transform these expressions into Mellin space. We define the Mellin amplitudes for the $\widetilde{\mathcal{F}}_{\pm}(U,V)$:
    \begin{equation}
        \begin{aligned}
            \widetilde{\mathcal{M}}_{\pm}(s,t) ={}& \frac{1}{2}\sum_{m,n=d-2}^\infty \frac{c^{(d)}_{m,n}}{s-2m}\left(\frac{1}{t-2n}\pm\frac{1}{\tilde u-2n}\right)\;,\\
            \widetilde{\mathcal{M}}_{\pm,\text{odd}}(s,t) ={}& \frac{1}{2}\sum_{m,n\,\in\,\frac{3(d-2)}{2}+\Z_{\geq0}} \frac{c^{(d),\text{odd}}_{m,n}}{s-2m}\left(\frac{1}{t-2n}\pm\frac{1}{\tilde u-2n}\right)\;.
        \end{aligned}
    \end{equation}
    In 3d, the $c$ coefficients are relatively simple:
    \begin{equation}
        \begin{aligned}
            c^{(3)}_{m,n} ={}& \frac{32\,\Gamma(m+1)\Gamma(n+1)\Gamma(m+n-\frac{1}{2})}{\pi^{7/2}\Gamma(m+\frac{1}{2})\Gamma(n+\frac{1}{2})\Gamma(m+n+1)}\;,\\
            c^{(3),\text{odd}}_{m,n} ={}& \frac{32\,\Gamma(m)\Gamma(n)\Gamma(m+n-\frac{1}{2})}{\pi^{7/2}\Gamma(m-\frac{1}{2})\Gamma(n-\frac{1}{2})\Gamma(m+n+1)}\;.
        \end{aligned}
    \end{equation}
    In 5d and 6d, the $c$ coefficients are more complex, and we present them in the attached Mathematica notebook. For reference, their lowest rows are
    \begin{gather*}
        c^{(5)}_{3,n} = \frac{2048\,(35n^2-5n-44)}{27\pi^4\,(n-1)n(n+1)}\;,\qquad c^{(6)}_{4,n} = \frac{90\,(5n^2-17n+9)}{7\,n(n-1)(n-2)}\;,\\
        c^{(5),\text{odd}}_{\frac92,n} = \frac{70\,\Gamma(n-2)\Gamma(n-1)\,(160n^2+416n+33)}{9\pi^3\,\Gamma(n-\frac72)\Gamma(n+\frac72)}\;,
    \end{gather*}
    with $c_{m,n}=c_{n,m}$.
    
     Considering the flavor indices, the reduced Mellin amplitudes at one loop are given by
    \begin{equation}
        \widetilde{\mathcal{M}}_{F|F}(s,t) = (\mathbf{d}_{st}{\mathcal{B}_{st}}+\mathbf{d}_{su}{\mathcal{B}_{su}}+\mathbf{d}_{tu}{\mathcal{B}_{tu}})\;,
    \end{equation}
    where
    \begin{equation}
        \begin{aligned}
            \mathcal{B}_{st}(s,t) ={}& \sum_{m,n=d-2}^\infty \frac{c_{m,n}^{(d)}}{(s-2m)(t-2n)} + \sum_{m,n\,\in\,\frac{3(d-2)}{2}+\Z_{\geq0}}\frac{c_{m,n}^{(d),\text{odd}}}{(s-2m)(t-2n)}\;,\\
            \mathcal{B}_{su}(s,t) ={}& \mathcal{B}_{st}(s,\tilde u)\;,\qquad \mathcal{B}_{tu}(s,t)=\mathcal{B}_{st}(t,\tilde u)\;,\qquad \tilde u=4d-12-s-t\;.
        \end{aligned}
    \end{equation}
    In 6d, the sum over the odd part is of course 0. In 5d and 6d the sums in $\mathcal{B}_{st}$ diverge, and the amplitude is defined only up to a polynomial in $s,t$ of the form of the contact terms of Section~\ref{sec:largeN}, which does not contribute to the double discontinuity.

	\subsection{CFT data from Lorentzian inversion}
	\label{subsec: inversion}
	
	We can now proceed to extract one-loop CFT data from our one-loop amplitudes, using the Lorentzian inversion formula. In this subsection we present dimension-agnostic formulae for various pieces of short and lowest twist long multiplet data, before evaluating these expressions in $d=3,5,6$ in the following subsection.
	
	It will be useful in what follows to define
	\begin{align}
  g_\beta(\bar{z}) = \bar{z}^{\frac{\beta}{2}}\, {}_2 F_{1}\!\left(\frac{\beta}{2}+1,\frac{\beta}{2};\beta;\bar{z}\right)\,, 
\end{align}
which is the leading lightcone block appropriate for the application of the Lorentzian inversion formula of \cite{Caron-Huot:2017vep} to the conformal block expansions \eqref{eq:6d_reduced_correlator}, \eqref{eq:5d_reduced_correlator} and \eqref{eq:3d_reduced_correlator}. In particular, the shift in the first argument of the hypergeometric function can be traced back to the external dimensions $(\Delta_{12},\Delta_{34})=(-2,0)$ appearing in the block decompositions, for which $g_\beta$ is the function $k_\beta$ of Section~\ref{sec:redMax}. In the notation of \cite{Caron-Huot:2017vep}, we have $(a,b)=(1,0)$. Accordingly, we have to be careful to include the factor of $(-1)^{a+b}=-1$ in the definition of the double discontinuity, such that in all that follows we have
\begin{align}
  \text{dDisc}[f(\bar{z})] = -\left(f(\bar{z}) - \frac{1}{2}f_{\circlearrowleft}(\bar{z}) - \frac{1}{2}f_{\circlearrowright }(\bar{z})\right)\,,
\end{align}
where $f_{\circlearrowleft}$ and $f_{\circlearrowright}$ are the analytic continuations of $f$ around $\bar{z}=1$ in the two directions. Finally we also define a normalization factor
\begin{align}
  \cN(\beta) = \frac{\Gamma(\frac{\beta}{2}-1)\Gamma(\frac{\beta}{2}+1)\Gamma(\frac{\beta}{2})^2}{4\pi^2\Gamma(\beta-1)\Gamma(\beta)} = \frac{\beta}{\beta-2}\,\frac{\Gamma(\frac{\beta}{2})^4}{4\pi^2\Gamma(\beta-1)\Gamma(\beta)}\,.
\end{align}
All of the CFT data we will extract can then be written in terms of a set of inversion integrals. For each $X\in \{st,tu\}$ we define
\begin{align}
  R^X_1(\ell)	&= 2\cN(2d+2\ell-2)\int_0^1 \frac{\text{d}\bar{z}}{\bar{z}^2} (1-\bar{z}) g_{2d+2\ell-2}(\bar{z}) \,\text{dDisc}\!\left[\left(\widetilde{\cG}_{F|F}(U,V) \Big|_{z^{d-2}\d_X}\right)(\bar{z})\right]\,,  		\nn\\
  R^X_2(\ell)	&= 16\cN(2d+2\ell-2)\int_0^1 \frac{\text{d}\bar{z}}{\bar{z}^2} (1-\bar{z}) g_{2d+2\ell-2}(\bar{z}) \,\text{dDisc}\!\left[\left(\widetilde{\cG}_{F|F}(U,V) \Big|_{z^{d-1}\log^2 z\,\d_X}\right)(\bar{z})\right]\,,  		\nn\\
  R^X_3(\ell)	&= 4\cN(2d+2\ell-2)\int_0^1 \frac{\text{d}\bar{z}}{\bar{z}^2} (1-\bar{z}) g_{2d+2\ell-2}(\bar{z}) \,\text{dDisc}\!\left[\left(\widetilde{\cG}_{F|F}(U,V) \Big|_{z^{d-1}\log z\,\d_X}\right)(\bar{z})\right]\,.  
\end{align}
The meaning of the notation inside the double discontinuity is that we expand the one-loop correlator around $z\to 0$, and extract the stated coefficient. Note that $R_2^{tu}$ vanishes identically, and thus there are five non-trivial inversion integrals to compute as a function of $\ell$. 

Note that these inversion integrals converge only above a threshold in $\ell$ that depends on $d$: $\ell\geq-1,0,1$ for $R_1^X$ and $\ell\geq0,1,2$ for $R_3^X$ in $d=3,5,6$, while $R_2^{st}$ converges for all $\ell\geq0$. Below these thresholds the double discontinuity does not fully fix the data obtained from $R_1^X$ and $R_3^X$, since in $d=5,6$ contact terms can contribute there, and continuing in $\ell$ does not necessarily recover them: in $d=6$ the continued integrals have poles at those spins, and in $d=5$ the finite continued values need not be the CFT data. 

We can then extract CFT data.
	We first consider the short multiplets $\cB[2]_\ell$, which appear in symmetric flavor channels for $\ell$ odd and antisymmetric ones for $\ell$ even. We can expand the squared OPE coefficients as
 \begin{align}
  \lambda_{\cB[2]_\ell}^2 = \dots + \frac{1}{c_J^2} (\lambda_{\cB[2]_\ell}^2)^{(2)} + \dots\,. 
\end{align}
Then, the one-loop contributions $(\lambda_{\cB[2]_\ell}^2)_\mathbf{r}^{(2)} $ for each flavor irrep $\r$ can be determined from
\begin{align}
  (\lambda_{\cB[2]_\ell}^2)^{(2)} = -\frac{8(d-4)(\ell+1)(d+\ell-2)}{d+2\ell-2} \Big[R_1^{st}(\ell)\big(\d_{st}-(-1)^\ell\d_{su}\big) + \big(1-(-1)^\ell\big)R_1^{tu}(\ell)\,\d_{tu}\Big]
  \label{eq: B2 OPE coefficient}
\end{align}
where we have suppressed flavor indices. From here we can also try to get the one-loop squared OPE coefficient of $\cD[4]$ by analytic continuation, as
 \begin{align}
  (\lambda_{\cD[4]}^2)^{(2)} = \lim_{\ell\to -1} \left[\frac{d-4}{2(\ell+1)}(\lambda_{\cB[2]_\ell}^2)^{(2)}\right]\,,
  \label{eq: D4 OPE coefficient}
\end{align}
where we set $(-1)^\ell=-1$ in \eqref{eq: B2 OPE coefficient}, since $\cD[4]$, like $\cB[2]_\ell$ with odd $\ell$, appears in symmetric flavor channels.
Note that in 3d the inversion integral is manifestly convergent down to $\ell=-1$ and thus this expression unambiguously fixes $(\lambda_{\cD[4]}^2)^{(2)}$, which is finite due to the simple pole of $\cN(\beta)$ at $\beta=2$. In contrast, in 5d and 6d the inversion integral diverges at $\ell=-1$. It can still be continued in $\ell$, as in \cite{Alday:2021ymb}, and we discuss below what this does and does not determine in each case.

We next turn to the long multiplets. For simplicity we focus on those of lowest twist, both because their data are most readily accessible, and also because there is a unique gluon double trace at lowest twist for each spin $\ell=0,1,2,\dots$. Note however that at $\ell=0$ there is a degeneracy with a graviton single trace in the flavor singlet channel, with mixing induced by a non-zero extremal coupling, as found for $d=4$ in \cite{Chester:2025wti} and for $d=3,6$ in \cite{Chester:2025jxg}. So we exclude the flavor singlet at $\ell=0$; together with the thresholds stated after the inversion integrals, the one-loop data of these multiplets are then determined for $\ell\geq0$ in $d=3$, $\ell\geq1$ in $d=5$ and $\ell\geq2$ in $d=6$.
 These multiplets then have scaling dimensions
\begin{align}
  (\Delta_{\ell})_\r =2(d-2)+\ell + \frac{1}{c_J} (\gamma_{\ell}^{(1)})_\r + \frac{1}{c_J^2}(\gamma_{\ell}^{(2)})_\r + \dots\,, 
\end{align}
where we have written only terms of order $(1/c_T)^0$, and the flavor irrep label $\r$ runs over symmetric representations for $\ell$ even and over antisymmetric ones for $\ell$ odd. Meanwhile their squared OPE coefficients can be expanded as
\begin{align}
  \lambda_\ell^2 = (\lambda_\ell^2)^{(0)} + \frac{1}{c_J}(\lambda_\ell^2)^{(1)} + \dots\,, 
\end{align}
which we recall carry suppressed flavor indices. The natural objects then appearing in the conformal block expansion, and thus in Lorentzian inversion formulae, are OPE coefficients weighted by anomalous dimensions. To retain our agnosticism towards the choice of flavor group $G_F$, it will be useful to use the compact notation
\begin{align}
  (\lambda_\ell^2)^{(i)} (\gamma_\ell^{(j)})^m = \sum_\r (\lambda_\ell^2)_\r^{(i)} ((\gamma_\ell^{(j)})_\r)^m P_\r
\end{align}
for any $m\in \N_0$. For $m=0$, this just recovers \eqref{eq: OPE coeff flavor expansion}.
 
We then first have that the one-loop correlator should reproduce tree-level data, as
\begin{align}
  (\lambda_\ell^2)^{(0)} (\gamma_\ell^{(1)})^2 = -4(d+2\ell)R_2^{st}(\ell)\big(\d_{st}+(-1)^\ell\d_{su}\big)\,. 
  \label{eq: tree from one-loop}
\end{align}
Note that this object is expressed in terms of the data we already provided in Section \ref{sec:gfft_tree_data} as\footnote{To derive this we used the identities $\mathbf{c}_t\circ\mathbf{c}_t=\d_{st}$, $\mathbf{c}_s\circ\mathbf{c}_t=\d_{su}-\d_{st}$, $\mathbf{c}_s\circ\mathbf{c}_s=2(\d_{st}-\d_{su})$, $\mathbf{t}_t\circ X=X$ and $\mathbf{t}_u\circ X=X^{BACD}$ (so that $\mathbf{t}_u\circ\d_{st}=\d_{su}$ and $\mathbf{t}_u\circ\d_{su}=\d_{st}$), which hold for any $G_F$. Here $\circ$ is the product defined below \eqref{eq: projector normalization}, $\mathbf{t}_t^{ABCD}\equiv\delta^{AD}\delta^{BC}=\sum_\r P_\r^{ABCD}$ and $\mathbf{t}_u^{ABCD}\equiv\delta^{AC}\delta^{BD}$. The data of Section~\ref{sec:gfft_tree_data} give $(\lambda_\ell^2)^{(0)}=A^{(d,0)}_{2,0,\ell}\big(\mathbf{t}_t+(-1)^\ell\mathbf{t}_u\big)$ and, after using $\mathbf{c}_u=-\mathbf{c}_s-\mathbf{c}_t$, $(\lambda_\ell^2)^{(0)}\gamma_\ell^{(1)}=2\,\omega^{(d)}_{2,0,\ell}\big(\mathbf{c}_s+(1+(-1)^\ell)\mathbf{c}_t\big)$. Since $\circ$ multiplies the irrep components, $(\lambda_\ell^2)^{(0)}\circ\big[(\lambda_\ell^2)^{(0)}(\gamma_\ell^{(1)})^2\big]=\big[(\lambda_\ell^2)^{(0)}\gamma_\ell^{(1)}\big]\circ\big[(\lambda_\ell^2)^{(0)}\gamma_\ell^{(1)}\big]$. The identities give $\big[(\lambda_\ell^2)^{(0)}\gamma_\ell^{(1)}\big]\circ\big[(\lambda_\ell^2)^{(0)}\gamma_\ell^{(1)}\big]=8\big(\omega^{(d)}_{2,0,\ell}\big)^2\big(\d_{st}+(-1)^\ell\d_{su}\big)$, and $(\lambda_\ell^2)^{(0)}\circ\big(\d_{st}+(-1)^\ell\d_{su}\big)=2A^{(d,0)}_{2,0,\ell}\big(\d_{st}+(-1)^\ell\d_{su}\big)$.}
\begin{align}
  (\lambda_\ell^2)^{(0)} (\gamma_\ell^{(1)})^2 = \frac{4\big(\omega^{(d)}_{2,0,\ell}\big)^2}{A^{(d,0)}_{2,0,\ell}}\big(\d_{st}+(-1)^\ell\d_{su}\big)\,,
\end{align}
and thus as a consistency check, we should find
\begin{align}
  R_2^{st}(\ell) &= -\frac{\big(\omega^{(d)}_{2,0,\ell}\big)^2}{(d+2\ell)\,A^{(d,0)}_{2,0,\ell}}\,.
\end{align}
We indeed find that this relation is satisfied in $d=3,5,6$ for all $\ell\geq0$, which is a non-trivial check on our Lorentzian inversion formulae. Since the integral for $R_2^{st}(\ell)$ converges for every $\ell\geq0$, no continuation in $\ell$ is needed here.

We can then finally move on to the new data. We find
\begin{align}
(\lambda_\ell^2)^{(0)}\gamma_\ell^{(2)} &=  -4(d+2\ell)\Big[R_3^{st}(\ell)\big(\d_{st}+(-1)^\ell\d_{su}\big) + \big(1+(-1)^\ell\big)R_3^{tu}(\ell)\,\d_{tu}\Big] \nn\\
	&\qquad + \frac{1}{2} \left(\partial_\ell + 1\right)\!\left[(\lambda_\ell^2)^{(0)} (\gamma_\ell^{(1)})^2\right]  - (\lambda_\ell^2)^{(1)} \gamma_\ell^{(1)}\,,
	  \label{eq: long multiplet anom dim}
\end{align}
where we emphasize that the derivative $\partial_\ell$ appearing here does \textit{not} act on the $(-1)^\ell$ factors appearing in \eqref{eq: tree from one-loop}, which are just there to impose the suitable even/odd spin projection, but it does act on the $(d+2\ell)$ prefactor. We note here that we need to input the tree-level combination $(\lambda_\ell^2)^{(1)} \gamma_\ell^{(1)}$, which is given in terms of the data of Section \ref{sec:gfft_tree_data} by\footnote{This uses the derivative relation $(\lambda_\ell^2)^{(1)}=\frac12\partial_n\big[(\lambda_\ell^2)^{(0)}\gamma_\ell^{(1)}\big]_{n=0}$ for the double traces of $\langle 2222\rangle$ at level $n$, continued in $n$. Since $(\lambda_\ell^2)^{(0)}\gamma_\ell^{(1)}=2\,\omega^{(d)}_{2,0,\ell}\big(\mathbf{c}_s+(1+(-1)^\ell)\mathbf{c}_t\big)$, it gives $(\lambda_\ell^2)^{(1)}=\partial_n\omega^{(d)}_{2,n,\ell}\big|_{n=0}\big(\mathbf{c}_s+(1+(-1)^\ell)\mathbf{c}_t\big)$, and then identities among flavor structures give the result. The derivative relation holds for contact interactions \cite{Heemskerk:2009pn,Fitzpatrick:2011dm} but fails for generic exchange diagrams \cite{Alday:2017gde}. We have verified it here directly in $d=3,5,6$ for $\ell=0,\dots,16$, by decomposing the tree-level amplitudes of Section~\ref{sec:gfft_tree_data}, and we assume it for higher spins. The continuation of $\omega^{(d)}_{2,n,\ell}$ in $n$ must be the one with at most polynomial growth, which is unique by Carlson's theorem. In $d=6$ this is the expression for $\omega^{(6)}_{p,n,\ell}$ in Section~\ref{sec:gfft_tree_data} as written. In $d=3,5$, the continuations needed are given in the attached Mathematica notebook.}
\begin{align}
  (\lambda_\ell^2)^{(1)} \gamma_\ell^{(1)} = \frac{\partial_n\big(\omega^{(d)}_{2,n,\ell}\big)^2\big|_{n=0}}{A^{(d,0)}_{2,0,\ell}}\big(\d_{st}+(-1)^\ell\d_{su}\big)\,.
\end{align}

	\subsection{Evaluation of CFT data}
	\label{subsec: CFT data evaluation}
	
We can now move on to evaluate the key formulae \eqref{eq: B2 OPE coefficient} and \eqref{eq: long multiplet anom dim} in each of $d=3,5,6$, and to see what \eqref{eq: D4 OPE coefficient} determines in each case. We write the one-loop data as
\begin{align}
  (\lambda^2_{\cB[2]_\ell})^{(2)} &= b_\ell\,\big(\d_{st}-(-1)^\ell\d_{su}\big)+\tilde b_\ell\,\d_{tu}\,,\qquad &&\tilde b_\ell=0\ \ \text{for even }\ell\,,\nn\\
  (\gamma^{(2)}_\ell)_\r &= g_\ell\,\big(\d_{st}+(-1)^\ell\d_{su}\big)_\r+\tilde g_\ell\,(\d_{tu})_\r\,,\qquad &&\tilde g_\ell=0\ \ \text{for odd }\ell\,,
  \label{eq: data parametrisation}
\end{align}
where $\r$ is any irrep in which the lowest twist long multiplet of spin $\ell$ appears, other than the singlet at $\ell=0$, which we excluded above. To extract $g_\ell,\tilde{g}_\ell$ from \eqref{eq: long multiplet anom dim} we need only note that 
\begin{align}
  (\lambda_\ell^2)^{(0)\,ABCD}&=A^{(d,0)}_{2,0,\ell}\big(\delta^{AD}\delta^{BC}+(-1)^\ell\,\delta^{AC}\delta^{BD}\big)\nn\\
  &=A^{(d,0)}_{2,0,\ell}\sum_\r\big(1+(-1)^\ell\epsilon_\r\big)P_\r^{ABCD}\,,\qquad P_\r^{BACD}=\epsilon_\r P_\r^{ABCD}\,,
\end{align}
from which it follows that GFFT OPE coefficients are equal in every flavor irrep in which they appear, i.e.\ 
$(\lambda^2_\ell)^{(0)}_\r=2A^{(d,0)}_{2,0,\ell}$. Hence $g_\ell$ and $\tilde g_\ell$ are the corresponding coefficients of $(\lambda_\ell^2)^{(0)}\gamma_\ell^{(2)}$ in \eqref{eq: long multiplet anom dim}, divided by $2A^{(d,0)}_{2,0,\ell}$. 

Below we then provide the data $b_\ell,\tilde{b}_\ell,g_\ell,\tilde{g}_\ell$ in $d=3,5,6$ in turn. Additionally in 3d we use \eqref{eq: D4 OPE coefficient} to determine $(\lambda_{\cD[4]}^2)^{(2)}$, which we match against localization, and in 5d and 6d we discuss what the continuation in $\ell$ determines.

\subsubsection*{CFT data in 3d}

We compute in closed form,
\begin{align}
  b_\ell	 			&=		\frac{32\,(\ell+1)\,\Gamma(\ell+2)^2}{3\pi^4(2\ell+1)\,\Gamma(2\ell+4)}\Big[\pi^2\big(16-(2\ell+3)\pi^2\big)+6\big((2\ell+3)\pi^2-4\big)H^{(2)}_{\ell+1}\Big]\,,\nn\\
  \tilde{b}_\ell	  	&=		\frac{64\,(\ell+1)\,\Gamma(\ell+2)^2}{3\pi^4\ell(\ell+3)(2\ell+1)\,\Gamma(2\ell+4)}\bigg[\pi^4\ell(\ell+3)(2\ell+3)-16\pi^2(\ell^2+3\ell+3)-12(2\ell+3)H^{(2)}_{\frac{\ell+1}{2}}\nn\\
  &\hspace{15mm}+6(\ell+3)\big(4(\ell+2)-\pi^2\ell(2\ell+3)\big)H^{(2)}_{\ell+1}+\frac{3\pi^4(2\ell+3)}{4^{\ell+1}}\binom{\ell+1}{\frac{\ell+1}{2}}^{\!2}\,\bigg]\qquad(\ell\ \text{odd})\,,\nn\\
  g_\ell				&= 		\frac{64\,(2\ell+3)^2}{\pi^4(\ell+1)^2(\ell+2)^2}\Big[3+\frac{2}{(\ell+2)(2\ell+3)}+4\big(H_{2\ell+2}-H_{\ell+1}\big)\Big]\nn\\
  &\hspace{15mm}-\frac{16\,(2\ell+3)\,Y_\ell}{\pi^4(\ell+1)(\ell+2)(2\ell+1)(2\ell+5)}\,,\nn\\
  \tilde{g}_\ell		&=		-\frac{32\,(2\ell+3)\,Y_\ell}{3\pi^4(\ell+1)(\ell+2)(2\ell+1)(2\ell+5)}\qquad(\ell\ \text{even})\,,
\end{align}
where
\begin{align}
  Y_\ell=36(2\ell+3)+\big((2\ell+3)^2+8\big)\pi^2-24(\ell+1)(\ell+2)\,H^{(2)}_{\ell+1}\,,
\end{align}
and $H_n=\sum_{j=1}^n\frac{1}{j}$ and $H^{(2)}_n=\sum_{j=1}^n\frac{1}{j^2}$ are harmonic numbers. For instance,
\begin{align}
  b_0&=\frac{16\,(34\pi^2-3\pi^4-24)}{9\pi^4}\,,&\tilde b_1&=\frac{4\,(95\pi^4-1048\pi^2+1200)}{135\pi^4}\,,\nn\\
  g_0&=\frac{480}{\pi^4}-\frac{408}{5\pi^2}\,,&\tilde g_0&=-\frac{192}{\pi^4}-\frac{272}{5\pi^2}\,.
\end{align}
Using \eqref{eq: D4 OPE coefficient} we find
\begin{align}
  (\lambda_{\cD[4]}^2)^{(2)}	=\frac{16}{3}\left(\frac{16}{\pi^2}-1\right)\big(\d_{st}+\d_{su}+\d_{tu}\big)\,.
  \label{eq: D4 3d}
\end{align}
The coefficient of $\d_{st}+\d_{su}$ comes from $R_1^{st}$ and that of $\d_{tu}$ from $R_1^{tu}$. That they agree, so that $(\lambda_{\cD[4]}^2)^{(2)}$ is totally symmetric in its flavor indices, is required by the crossing relation of Section~\ref{sec:red3Half}, since $\lambda^2_{\cD[2]}\propto1/c_J$ exactly.

Let us then compare this against results from supersymmetric localization. In $d=3$, $\cD[2]$ and $\cD[4]$ contribute to the topological sector of Section~\ref{sec:red3Half} \cite{Chester:2014mea,Beem:2016cbd}, whose correlators can be computed from the $S^3$ partition function $Z$ deformed by real masses for $G_F$ \cite{Dedushenko:2016jxl}. We take the 3d $\mathcal{N}=4$ $U(N)$ gauge theory with one adjoint and $k\geq2$ fundamental hypermultiplets, which at large $N$ is dual to M-theory on AdS$_4\times S^7/\mathbb{Z}_k$, where the $G_F=SU(k)$ gluons live on the fixed locus AdS$_4\times S^3$ \cite{Chester:2023qwo}. We turn on a real mass $m^A=m\,n^A$ along the Cartan subalgebra, with $n^AT^A=\mathrm{diag}(t_1,\dots,t_k)$ in the fundamental representation. With $\psi^2=2$ we have $\mathrm{Tr}_{\text{fund}}(T^AT^B)=\delta^{AB}$, so that $n\cdot n=\sum_It_I^2$. Writing $X^{nnnn}\equiv X^{ABCD}n^An^Bn^Cn^D$, the topological sector gives
\begin{align}
  \partial_m^2\log Z\Big|_{m=0}=-\frac{\pi^2}{8}\,c_J\,n\cdot n\,,\qquad
  \frac{\partial_m^4\log Z}{\big(\partial_m^2\log Z\big)^2}\bigg|_{m=0}=\frac{(\lambda^2_{\cD[4]})^{nnnn}}{(n\cdot n)^2}-2\,,
  \label{eq: 1d relations}
\end{align}
exactly in $1/c_J$ \cite{Closset:2012vg,Dedushenko:2016jxl,Chester:2023qwo}. In the second relation $\lambda^2_{\cD[2]}$ drops out, being antisymmetric in its first two indices, and the right-hand side vanishes for generalized free fields, for which $(\lambda^2_{\cD[4]})^{(0)\,nnnn}=2(n\cdot n)^2$. At large $N$, $Z=e^{A}C^{-\frac13}\mathrm{Ai}\big[C^{-\frac13}(N-B)\big]$ up to non-perturbative corrections \cite{Mezei:2013gqa,Grassi:2014vwa,Chester:2023qwo}, where $C=\frac{2}{\pi^2k}$ does not depend on $m$, $B$ is quadratic in $m$, and \cite{Chester:2023qwo,Kubo:2025dot}
\begin{align}
  \partial_m^4A\Big|_{m=0}=16\,\mathfrak{b}\sum_{I,J=1}^k(t_I-t_J)^4=\frac{16\,\mathfrak{b}}{3}\,\big(\d_{st}+\d_{su}+\d_{tu}\big)^{nnnn}\,,\qquad \mathfrak{b}=\frac{\pi^2}{4}-\frac{\pi^4}{64}\,.
  \label{eq: d4A}
\end{align}
In the second equality we used $\d_X^{nnnn}=\mathrm{Tr}\,(\mathrm{ad}\,T)^4=\sum_{I,J}(t_I-t_J)^4$ for each $X\in\{st,su,tu\}$, where $T=n^AT^A$.\footnote{For $T=T^1\equiv\frac12\mathrm{diag}(1,-1,0,\dots,0)$, and for $T=T^3\equiv\frac{1}{2\sqrt6}\mathrm{diag}(1,1,1,-3,0,\dots,0)$ when $k\geq4$, both with $\mathrm{Tr}\,T^2=\frac12$, \eqref{eq: d4A} gives $4(6+k)\mathfrak{b}$ and $4\big(6+\frac76k\big)\mathfrak{b}$. These are the values in (5.18) of \cite{Chester:2023qwo} and in (4.12) of \cite{Kubo:2025dot}, where $q$ is our $k$. Since $\mathrm{Tr}\,T^4/(\mathrm{Tr}\,T^2)^2$ differs between $T^1$ and $T^3$, these two values fix $\partial_m^4A$ for every $n$; for $k=2,3$ there is a single quartic invariant and $T^1$ suffices.}
Since $\partial_m^4B=0$, the terms involving $B$ contribute to $\partial_m^4\log Z|_{m=0}$ only at order $N^{-\frac12}$. The first relation in \eqref{eq: 1d relations} gives $c_J=\frac{16\sqrt{2kN}}{\pi}-2k\big(1+\frac{4}{\pi^2}\big)+O(N^{-\frac12})$.\footnote{In the normalization \eqref{charges}, a free hypermultiplet in the fundamental of $SU(k)$ has $c_J=8$. The expression printed for $c_J$ in (5.20) of \cite{Chester:2023qwo} is half of the one given here, although $c_J$ is defined there as in \eqref{charges}; the constant $\mathfrak{a}$ of (5.21) there equals $\frac14+\frac{\pi^2}{16}$, as follows from (3.63) of \cite{Kubo:2025dot}.}

We now expand \eqref{eq: 1d relations} in $1/c_J$. At order $1/c_J$, $(\lambda^2_{\cD[4]})^{(1)}$, like the rest of the tree-level correlator, is a combination of $\mathbf{c}_s$, $\mathbf{c}_t$ and $\mathbf{c}_u$, which vanish on $n^An^Bn^Cn^D$. Graviton exchange, of order $1/c_T\sim k/c_J^3$, and the $O(N^{-\frac12})$ terms above enter only at order $c_J^{-3}$. Nor does any contact term contribute at order $c_J^{-2}$: the one-loop amplitude is UV finite in $d=3$, and the first contact terms, of the form $(\mathrm{Tr}\,F^2)^2$ and $\mathrm{Tr}\,F^4$, would appear at order $N^{-\frac76}\sim c_J^{-\frac73}$, where in fact they vanish \cite{Chester:2023qwo}. Hence
\begin{align}
  (\lambda^2_{\cD[4]})^{(2)\,nnnn}=\frac{64}{\pi^4}\,\partial_m^4A\Big|_{m=0}=\frac{1024\,\mathfrak{b}}{3\pi^4}\,\big(\d_{st}+\d_{su}+\d_{tu}\big)^{nnnn}\,,
\end{align}
in exact agreement with \eqref{eq: D4 3d} for every $k\geq2$ and every Cartan direction $n$, since $\frac{1024\,\mathfrak{b}}{3\pi^4}=\frac{16}{3}\big(\frac{16}{\pi^2}-1\big)$. Both sides are totally symmetric invariant tensors, which are fixed by their values on the Cartan subalgebra, so they agree as tensors. For $k\geq4$ this tests both independent quartic invariants, $(\mathrm{Tr}\,T^2)^2$ and $\mathrm{Tr}\,T^4$.

\subsubsection*{CFT data in 5d}

In $d=5$, at every integer spin above the thresholds we find
\begin{align}
  b_\ell &= a^{(5)}_1(\ell)+a^{(5)}_2(\ell)\,\frac{\sqrt3}{\pi}+\frac{a^{(5)}_3(\ell)}{\pi^2}+a^{(5)}_4(\ell)\,\frac{\sqrt3}{\pi^3}+\frac{a^{(5)}_5(\ell)}{\pi^4}\,, &&\ell\geq0\,,\nn\\
  \tilde b_\ell &= a^{(5)}_6(\ell)+a^{(5)}_7(\ell)\,\frac{\sqrt3}{\pi}+\frac{a^{(5)}_8(\ell)}{\pi^2}+a^{(5)}_9(\ell)\,\frac{\sqrt3}{\pi^3}+\frac{a^{(5)}_{10}(\ell)}{\pi^4}\,, &&\ell\geq1\ \text{odd}\,,\nn\\
  g_\ell &= a^{(5)}_{11}(\ell)\,\frac{\sqrt3}{\pi}+\frac{a^{(5)}_{12}(\ell)}{\pi^2}+a^{(5)}_{13}(\ell)\,\frac{\sqrt3}{\pi^3}+\frac{a^{(5)}_{14}(\ell)}{\pi^4}\,, &&\ell\geq1\,,\nn\\
  \tilde g_\ell &= a^{(5)}_{15}(\ell)\,\frac{\sqrt3}{\pi}+\frac{a^{(5)}_{16}(\ell)}{\pi^2}+a^{(5)}_{17}(\ell)\,\frac{\sqrt3}{\pi^3}+\frac{a^{(5)}_{18}(\ell)}{\pi^4}\,, &&\ell\geq2\ \text{even}\,,
  \label{eq: 5d data}
\end{align}
where each $a^{(5)}_i(\ell)$ is a rational number for every integer $\ell$ in the range indicated. We provide all eighteen functions $a^{(5)}_i(\ell)$ in the attached Mathematica notebook. For instance at the lowest spins,
\begin{align}
  b_0 &= \tfrac{1964288}{59049} - \tfrac{131292725248}{1594323}\,\tfrac{\sqrt3}{\pi} + \tfrac{4719817041376}{14073345}\,\tfrac{1}{\pi^2} \nn\\
  &\quad + \tfrac{123649024}{405}\,\tfrac{\sqrt3}{\pi^3} - \tfrac{6037735936}{10935}\,\tfrac{1}{\pi^4}\,, \nn\\
  \tilde b_1 &= \tfrac{844050248}{2066715} + \tfrac{3680289947648}{2657205}\,\tfrac{\sqrt3}{\pi} - \tfrac{86648308281344}{15324309}\,\tfrac{1}{\pi^2} \nn\\
  &\quad - \tfrac{363665293312}{70875}\,\tfrac{\sqrt3}{\pi^3} + \tfrac{9891637889024}{1063125}\,\tfrac{1}{\pi^4}\,, \nn\\
  g_1 &= -\tfrac{364214076047360}{4782969}\,\tfrac{\sqrt3}{\pi} + \tfrac{450575655177664}{1449981}\,\tfrac{1}{\pi^2} \nn\\
  &\quad + \tfrac{7191368966144}{25515}\,\tfrac{\sqrt3}{\pi^3} - \tfrac{36224202674176}{70875}\,\tfrac{1}{\pi^4}\,, \nn\\
  \tilde g_2 &= \tfrac{766761276276736}{531441}\,\tfrac{\sqrt3}{\pi} - \tfrac{320269445967712672}{54392877}\,\tfrac{1}{\pi^2} \nn\\
  &\quad - \tfrac{75694199566336}{14175}\,\tfrac{\sqrt3}{\pi^3} + \tfrac{4805142623577088}{496125}\,\tfrac{1}{\pi^4}\,.
\end{align}
In $d=5$ the continuation of $R_1^X(\ell)$ is regular at $\ell=-1$, and the limit in \eqref{eq: D4 OPE coefficient} gives
\begin{align}
  \lim_{\ell\to -1} \left[\frac{1}{2(\ell+1)}(\lambda_{\cB[2]_\ell}^2)^{(2)}\right] &= D_{st}\,\big(\d_{st}+\d_{su}\big)+D_{tu}\,\d_{tu}\,,\nn\\
  D_{st} &= \tfrac{1964288}{59049} + \tfrac{13576437760}{1594323}\,\tfrac{\sqrt3}{\pi} - \tfrac{88432909088}{2525985}\,\tfrac{1}{\pi^2} \nn\\
  &\quad - \tfrac{22770688}{729}\,\tfrac{\sqrt3}{\pi^3} + \tfrac{13485568}{243}\,\tfrac{1}{\pi^4}\approx 5.7848\,, \nn\\
  D_{tu} &= -\tfrac{1419968}{6561} + \tfrac{14415822848}{531441}\,\tfrac{\sqrt3}{\pi} - \tfrac{91078968128}{841995}\,\tfrac{1}{\pi^2} \nn\\
  &\quad - \tfrac{73154560}{729}\,\tfrac{\sqrt3}{\pi^3} + \tfrac{43561984}{243}\,\tfrac{1}{\pi^4}\approx 13.6381\,.
  \label{eq: D4 5d}
\end{align}
The coefficient $(\lambda^2_{\cD[4]})^{(2)}$ equals this up to the contribution of the contact term $\widetilde{\cG}'_{F^4}$ of Section~\ref{sec:largeN}, a totally symmetric tensor that our analysis does not fix.\footnote{The limit \eqref{eq: D4 5d} is the value of $(\lambda^2_{\cD[4]})^{(2)}$ for the amplitude obtained by cutting off the sums in $\mathcal{B}_{st}$, $\mathcal{B}_{su}$ and $\mathcal{B}_{tu}$ at $m+n\leq n_{\max}$ and dropping the divergence, which is proportional to $n_{\max}^{1/2}\,(\d_{st}+\d_{su}+\d_{tu})$. The full amplitude at this order can differ from this one only by a constant, which crossing requires to be totally symmetric.} Hence only its part that is not totally symmetric is determined: $\frac13(D_{st}-D_{tu})\,\big(\d_{st}+\d_{su}-2\,\d_{tu}\big)$, with $\frac13(D_{st}-D_{tu})\approx-2.6178$. Equating \eqref{eq: D4 5d} with $(\lambda^2_{\cD[4]})^{(2)}$ is the analogue of the conjecture made for ABJ(M) theory in \cite{Alday:2021ymb}, that continuing the inversion formula below its range of convergence fixes such contact terms. That conjecture was verified using localization for the one-loop graviton term $R|R$ \cite{Alday:2021ymb,Alday:2022rly}, whereas its analogue \cite{Aprile:2017bgs} was already known to fail in 4d $\mathcal{N}=4$ super Yang-Mills theory \cite{Alday:2017vkk,Chester:2019pvm}; in $d=5$ we know of no such test.

\subsubsection*{CFT data in 6d}

In $d=6$, at every integer spin above the thresholds we find
\begin{align}
  b_\ell &= a^{(6)}_1(\ell)+a^{(6)}_2(\ell)\,\log 2+a^{(6)}_3(\ell)\,\zeta(3)\,, &&\ell\geq1\,,\nn\\
  \tilde b_\ell &= a^{(6)}_4(\ell)+a^{(6)}_5(\ell)\,\log 2+a^{(6)}_6(\ell)\,\zeta(3)\,, &&\ell\geq1\ \text{odd}\,,\nn\\
  g_\ell &= a^{(6)}_7(\ell)+a^{(6)}_8(\ell)\,\log 2+a^{(6)}_9(\ell)\,\zeta(3)\,, &&\ell\geq2\,,\nn\\
  \tilde g_\ell &= a^{(6)}_{10}(\ell)+a^{(6)}_{11}(\ell)\,\log 2+a^{(6)}_{12}(\ell)\,\zeta(3)\,, &&\ell\geq2\ \text{even}\,,
  \label{eq: 6d data}
\end{align}
where each $a^{(6)}_i(\ell)$ is a rational number for every integer $\ell$ in the range indicated.
 We provide all twelve functions $a^{(6)}_i(\ell)$ in the attached Mathematica notebook. For instance at the lowest spins,
\begin{align}
  b_1 &= \tfrac{63446713825}{1806336} - \tfrac{2764000}{21}\log 2 + \tfrac{382353125}{8192}\,\zeta(3)\,, \nn\\
  \tilde b_1 &= \tfrac{1549822045}{18816} - \tfrac{19445600}{63}\log 2 + \tfrac{28021875}{256}\,\zeta(3)\,, \nn\\
  g_2 &= \tfrac{6386964331127}{1267200} - 18886560\log 2 + \tfrac{6858372465}{1024}\,\zeta(3)\,, \nn\\
  \tilde g_2 &= \tfrac{1731143604919}{147840} - \tfrac{219388704}{5}\log 2 + \tfrac{3983387625}{256}\,\zeta(3)\,.
\end{align}
Unlike in $d=3$, \eqref{eq: D4 OPE coefficient} does not determine $(\lambda^2_{\cD[4]})^{(2)}$ in $d=6$. The inversion integrals can be continued in $\ell$, but the continuation has simple poles at the physical spins below the thresholds.
 In particular, as $\ell\to-1$ we find
\begin{align}
  \frac{1}{\ell+1}\,(\lambda^2_{\cB[2]_\ell})^{(2)}=-\frac{135}{14}\,\frac{\d_{st}+\d_{su}-2\,\d_{tu}}{\ell+1}+O(1)\,.
  \label{eq: 6d D4 pole}
\end{align}
This pole, and the analogous poles at the other spins below the thresholds, are the imprint of the logarithmic divergence of the one-loop amplitude.
 Cutting off the sums over $m,n$ in $\mathcal{B}_{st}$ at $m+n\leq n_{\max}$, the reduced Mellin amplitude diverges logarithmically as
\begin{align}
  \widetilde{\mathcal{M}}_{F|F}\supset\frac{75}{64}\,\log n_{\max}\,\Big[\d_{st}\,(s+t-8)+\d_{su}\,(s+\tilde u-8)+\d_{tu}\,(t+\tilde u-8)\Big]\,.
  \label{eq: 6d log divergence}
\end{align}
There is in addition a power divergence proportional to $n_{\max}(\d_{st}+\d_{su}+\d_{tu})$, which is absorbed by the $F^4$ contact term. The logarithmic divergence is canceled by a $D^2F^4$ counterterm whose finite part is not fixed by our analysis, and so the $O(1)$ term in \eqref{eq: 6d D4 pole} does not determine $(\lambda^2_{\cD[4]})^{(2)}$. These poles are the same kind of obstruction to the continuation in spin as was found for the $R|R^4$ term in \cite{Alday:2021ymb}.
 What is fixed is the term $\widetilde{\cG}_{\log}$: since the cutoff scales\footnote{The Mellin variables are flat space Mandelstam invariants in units of the AdS radius $L$, so that $n_{\max}\sim L^2\Lambda^2$ with $\Lambda$ the scale at which the ten-dimensional gauge theory is UV completed; for the E-string theory \cite{Alday:2021odx} $\Lambda\sim\ell_{11}^{-1}$ and $c_J\propto(L/\ell_{11})^6$, with $\ell_{11}$ the eleven-dimensional Planck length.} as $n_{\max}\propto c_J^{1/3}$, its reduced Mellin amplitude is $\frac13$ of the coefficient of $\log n_{\max}$ in \eqref{eq: 6d log divergence}, and for instance the coefficient of $\frac{\log c_J}{c_J^2}$ in $\lambda^2_{\cD[4]}$, which we denote by $(\lambda^2_{\cD[4]})^{(\log)}$, is $\frac13$ of the residue in \eqref{eq: 6d D4 pole},
\begin{align}
  (\lambda^2_{\cD[4]})^{(\log)}=-\frac{45}{14}\,\big(\d_{st}+\d_{su}-2\,\d_{tu}\big)\,.
\end{align}
\subsection{Flat space limit}\label{subsec: flat space}

Finally, we check the flat space limit of our one-loop amplitude, as was done for $d=4$ in \cite{Alday:2021ajh,Behan:2023fqq}. The gluons propagate on the brane worldvolume AdS$_{d+1}\times S^3$ \cite{Aharony:1998xz,Alday:2021odx}, where AdS$_{d+1}$ has radius $L$ and $S^3$ has radius $\frac{2L}{d-2}$,\footnote{In $d=5$ the bulk geometry is a warped product, and $L$ should then be understood as the effective radius seen by the gauge field on the brane. Only the ratio of the two radii enters our final results.} and so at energies much larger than $1/L$ they are described by maximally supersymmetric Yang-Mills theory in $d+4$ flat dimensions. We normalize its coupling $g_\text{YM}$ such that the action is $-\frac{1}{2g_\text{YM}^2}\int \text{d}^{d+4}x\, F^A_{MN}F^{A\,MN}+\dots$, with the structure constants normalized as in Section~\ref{sec:largeN}. After reduction on $S^3$, the current two-point function computed from a gauge field on AdS$_{d+1}$ \cite{Freedman:1998tz} relates this coupling to the flavor central charge defined in \eqref{charges}, as
\begin{align}
  g_\text{YM}^2 = \frac{64\pi^{\frac{d+4}{2}}\,\Gamma(d)}{(d-2)^2\,\Gamma(\frac{d}{2})^3}\,\frac{L^d}{c_J}\,.
  \label{eq: gYM cJ}
\end{align}
At large $s,t$ with $s/t$ fixed, the Mellin amplitude defined in \eqref{mellinDef} is determined by the flat space scattering amplitude \cite{Penedones:2010ue,Fitzpatrick:2011hu,Fitzpatrick:2011dm}, as
\begin{align}
  \cM(s,t;\alpha) \approx \frac{L^{3-d}}{8\pi^{\frac{d}{2}}\,\Gamma(\frac{d}{2}-1)^2\,\Gamma(d-2)^2\,\text{Vol}(S^3)}\int_0^\infty \text{d}\beta\,\beta^{\frac{3d}{2}-5}e^{-\beta}\,\cA\Big(\frac{2\beta s}{L^2},\frac{2\beta t}{L^2};\alpha\Big)\,,
  \label{eq: flat limit formula}
\end{align}
where $\text{Vol}(S^3)=\frac{16\pi^2L^3}{(d-2)^3}$ arises from the reduction on $S^3$, and $\cA(s,t;\alpha)$ is the amplitude for four gluons with momenta $p_i$ in $\R^{1,d}$ and polarizations $e_i$ along the three remaining directions, which descend from $S^3$. It is a function of the Mandelstam variables $s=-(p_1+p_2)^2$, $t=-(p_1+p_4)^2$ and $u=-(p_1+p_3)^2=-s-t$. We take all momenta to be incoming and define $\cA$ such that the $S$-matrix is $1+i(2\pi)^{d+4}\delta^{(d+4)}(p_1+p_2+p_3+p_4)\,\cA$. As in \cite{Alday:2021odx}, the polarizations are null vectors with $e_i\cdot e_j=(t_i\cdot t_j)^2$ in terms of the $t_i$ of Section~\ref{sec:fullHalf}, and we have stripped off a factor of $(e_1\cdot e_2)(e_3\cdot e_4)$, so that $\cA$ depends on them only through $\alpha$, since $\frac{(e_1\cdot e_3)(e_2\cdot e_4)}{(e_1\cdot e_2)(e_3\cdot e_4)}=\alpha^2$ and $\frac{(e_1\cdot e_4)(e_2\cdot e_3)}{(e_1\cdot e_2)(e_3\cdot e_4)}=(1-\alpha)^2$. At tree level and at one loop these amplitudes are \cite{Green:1982sw,Bern:1998ug}
\begin{align}
  \cA_\text{tree} &= -g_\text{YM}^2\,(u+s\alpha)^2\Big(\frac{\mathbf{c}_s}{su}-\frac{\mathbf{c}_t}{tu}\Big)\,,		\nn\\
  \cA_\text{1-loop} &= \frac{g_\text{YM}^4}{(4\pi)^{\frac{d+4}{2}}}\,(u+s\alpha)^2\Big(\d_{st}\,I_{d+4}(s,t)+\d_{su}\,I_{d+4}(s,u)+\d_{tu}\,I_{d+4}(t,u)\Big)\,,
  \label{eq: flat amplitudes}
\end{align}
with the flavor structures of Sections~\ref{sec:gfft_tree_data} and~\ref{sec:AnswerMel}. Here $I_D$ is the scalar box integral in $D$ dimensions, which for $4<D<8$ and $s,t<0$ is given by
\begin{align}
  I_D(s,t) &= \Gamma\big(4-\tfrac{D}{2}\big)\int_0^1 \text{d}^4a\,\,\delta\Big(1-\sum_{i=1}^4 a_i\Big)\bigl(-s\,a_1a_3-t\,a_2a_4\bigr)^{\frac{D}{2}-4}	\nn\\
  &= \frac{\Gamma(\frac{D-4}{2})}{\Gamma(D-4)}\int_0^\infty \text{d}S\,\text{d}T\,\frac{(ST)^{\frac{D-6}{2}}}{(S+T)^{\frac{D-4}{2}}}\,\frac{1}{(S-s)(T-t)}\,,
  \label{eq: box}
\end{align}
and elsewhere by analytic continuation in $s,t$. For $D\geq 8$ the box is ultraviolet divergent and depends on the regulator through a polynomial in $s,t$, but the weight multiplying $\frac{1}{(S-s)(T-t)}$ in the second line remains its double spectral density.

As a check of these normalizations, inserting $\cA_\text{tree}$ and \eqref{eq: gYM cJ} into \eqref{eq: flat limit formula} gives
\begin{align}
  \cM_{F,2}(s,t;\alpha) \to -\tau_d\,(u+s\alpha)^2\Big(\frac{\mathbf{c}_s}{su}-\frac{\mathbf{c}_t}{tu}\Big)\,,\qquad \tau_d=\frac{(d-1)^2(d-2)^5\,\Gamma(\frac{3d}{2}-4)}{8\,\Gamma(d)\,\Gamma(\frac{d}{2})^5}\,,
  \label{eq: flat tree}
\end{align}
so that $\tau_d=\frac{8}{\pi^2},\frac{160}{\pi^2},20$ for $d=3,5,6$. This is indeed the behavior of the tree-level amplitude \eqref{eq:gluon_tree_mellin} at large $s,t$. Note that at fixed $c_J$ we have $g_\text{YM}^2\propto\text{Vol}(S^3)$, so that $\text{Vol}(S^3)$ cancels in $\tau_d$ but not in the one-loop coefficient below, and thus only the latter is sensitive to the radius of $S^3$.

At one loop, we also need the relation $\cM=\mathcal{D}_d\widetilde{\cM}$ between the full and reduced Mellin amplitudes \cite{Chester:2025jxg}. At large $s,t$ it reduces to $\cM\to 2(u+s\alpha)^2\big(\frac{3(d-2)}{2}+s\partial_s+t\partial_t\big)\widetilde{\cM}$, which on the boxes, of degree $\frac{d-4}{2}$ in $s,t$, is multiplication by $2(2d-5)(u+s\alpha)^2$. Then \eqref{eq: flat limit formula} predicts that at large $s,t$,
\begin{align}
  \widetilde{\cM}_{F|F}(s,t) &\to \eta_d\,\Big(\d_{st}\,I_{d+4}(s,t)+\d_{su}\,I_{d+4}(s,u)+\d_{tu}\,I_{d+4}(t,u)\Big)\,,		\nn\\
  \eta_d &= \frac{(d-1)^2(d-2)^3\,\Gamma(2d-5)}{2^{\frac{d}{2}+2}\,\Gamma(\frac{d}{2})^8}\,,
  \label{eq: flat prediction}
\end{align}
so that $\eta_d=\frac{64\sqrt2}{\pi^4},\frac{262144\sqrt2}{81\pi^4},\frac{1125}{8}$ for $d=3,5,6$.

To compare this with our result, note that at large $s,t$ the sums in $\mathcal{B}_{st}$ are dominated by large $m,n$. At large $m,n$ with $m/n$ fixed, we find
\begin{align}
  c^{(d)}_{m,n}\ \to\ C_d\,\frac{(mn)^{\frac{d-2}{2}}}{(m+n)^{\frac{d}{2}}}\,,\qquad C_3=\frac{32}{\pi^{7/2}}\,,\quad C_5=\frac{32768}{81\pi^{7/2}}\,,\quad C_6=\frac{75}{4}\,,
  \label{eq: cmn asymptotics}
\end{align}
and the same for $c^{(d),\text{odd}}_{m,n}$ in $d=3,5$. This has the form of the double spectral density in \eqref{eq: box} at $D=d+4$, with $S=2m$ and $T=2n$. Replacing the sums over $m,n$ by integrals, as in \cite{Alday:2021ajh}, we thus find
\begin{align}
  \mathcal{B}_{st}(s,t) \to \frac{(2-\delta_{d,6})\,C_d\,\Gamma(d)}{2^{\frac{d}{2}}\,\Gamma(\frac{d}{2})}\,I_{d+4}(s,t)= \eta_d\, I_{d+4}(s,t)\,,
  \label{eq: Bst flat}
\end{align}
where the factor of $2$ in $d=3,5$ is due to the odd twist double traces, which contribute the same as the even twist ones in the flat space limit, as for gravitons in $d=3$ \cite{Alday:2022rly}. Since $\tilde u\approx u$ at large $s,t$, the terms $\mathcal{B}_{su}$ and $\mathcal{B}_{tu}$ similarly reproduce the other two boxes, and so our one-loop amplitude matches the flat space prediction \eqref{eq: flat prediction} in each of $d=3,5,6$.

In $d=3$ both the sums in $\mathcal{B}_{st}$ and the box converge, and \eqref{eq: flat prediction} holds as written for $s$, $t$ and $u$ away from the positive real axis, which we have also verified numerically for the full function $\mathcal{B}_{st}(s,t)$ along several directions in the complex $s,t$ planes. In $d=5,6$ the nine- and ten-dimensional boxes are ultraviolet divergent, as are the sums in $\mathcal{B}_{st}$. What we have matched is then the double spectral density, which fixes the amplitude up to a constant in $d=5$ and a polynomial of degree at most one in $s,t$ in $d=6$. These are the reduced Mellin amplitudes of the contact terms $\widetilde{\cG}'_{F^4}$ and $\widetilde{\cG}_{D^2F^4}$ of Section~\ref{sec:largeN}. With the same cutoff, the divergences also agree at large $s,t$. In $d=6$, cutting off the integral in \eqref{eq: box} at $S+T\leq 2n_{\max}$ gives $I_{10}(s,t)\supset\frac{s+t}{120}\log n_{\max}$, and $\frac{1}{120}\eta_6=\frac{75}{64}$ is precisely the coefficient of $(s+t)\log n_{\max}$ in \eqref{eq: 6d log divergence}, while in $d=5$ both sides diverge only as $n_{\max}^{1/2}$, with no logarithm; as in $d=6$, this power divergence is absorbed by the $F^4$ contact term $\widetilde{\cG}_{F^4}$.

\section{Conclusion}\label{sec:conclusion}

In this paper we found a new formal solution to the superconformal Ward identities for the stress tensor multiplet correlator in 3d maximally supersymmetric CFTs, and the flavor multiplet correlator in $d=3,5,6$ half-maximally supersymmetric CFTs. Our solution resolves the problem of non-locality in the old solution of \cite{Dolan:2004mu} for odd $d$; in $d=6$ it is equivalent to that solution. We show that our reduced correlator obeys simple crossing equations, up to elements of the kernel of the map to the full correlator, and has a superblock expansion with just one block per supermultiplet. In 3d, we find that regularity of the reduced superblock expansion under the topological twist $z=\bar z=1/\alpha$ requires an additional term in the solution to the Ward identity, which corresponds to the 1d topological sector. We then used our new reduced correlators to compute the one-loop correction to gluon scattering in $d=3,5,6$, which we matched to the flat space limit, and to protected CFT data in 3d that follow from supersymmetric localization.

Our new reduced correlators should facilitate numerical bootstrap studies of these theories. In particular, all previous numerical bootstrap studies \cite{Chester:2014fya,Chester:2014mea,Agmon:2019imm,Agmon:2017xes,Alday:2021ymb,Alday:2022rly,Chester:2024bij} of the stress tensor correlator in maximally supersymmetric 3d CFTs used the full correlator. It was observed in \cite{Chester:2014fya} that supersymmetry reduced the original six crossing equations to one crossing equation, as well as derivatives in $z$ (but not $\bar z$) of a second crossing equation. This was in contrast to the stress tensor correlator in maximally supersymmetric CFTs in $d=4,6$, where there was just one independent crossing equation, which followed from the known reduced correlators in those cases. Our new reduced correlator implies that we will just have one crossing equation too if we work with the superblock expansion of the reduced correlator, up to a multiple of the kernel element $U(U-V)$ and together with the relation (\ref{eq:crossing_twisted_max}) among the protected OPE coefficients. It will be interesting to see if the different number of independent crossing equations in the full versus reduced correlators leads to more constraining bounds in one case.

It would also be interesting to generalize our reduced correlators to the most general four-point function of half-BPS operators in both the maximal and half-maximal cases. In 4d \cite{Bissi:2015qoa,Doobary:2015gia}, it was found that simple superblock expansions exist for all such correlators, and it would be nice if this generalized to other $d$, beyond the next-to-next-to-extremal correlators treated in \cite{Woolley:2026cii,Woolley:2026anb}. This would also facilitate numerical bootstrap studies in those cases.

Finally, it would be nice to understand for which correlators in general reduced correlators exist. A prerequisite is that superconformal symmetry must impose constraints on the four-point function of the superprimary itself, and not just relate it to correlators of other operators in the same multiplet. So far this has been observed both for correlators of half-BPS operators, as discussed here, and for correlators of $1/3$-BPS operators in \cite{Binder:2019mpb,Binder:2020ckj}. In the latter case, these Ward identities did not impose all the constraints of supersymmetry on the superprimary correlator, for which one instead needed the stronger constraints coming from the superblock expansion, which was derived from the superconformal Casimir equation. Nonetheless, it is still possible that a reduced correlator could be defined in that and similar cases.

\section*{Acknowledgments} 
We thank Ofer Aharony and Silviu Pufu for useful discussions. SMC is supported by the Royal Society under the grant URF\textbackslash R1\textbackslash 221310 and the UK Engineering and Physical Sciences Research Council grant number EP/Z000106/1. RM is also supported by EP/Z000106/1.  The work of CV is supported by the Fonds de recherche du Qu\'ebec, secteur Nature et technologies, through its Doctoral Training Scholarships program, scholarship number 349150 (\href{https://doi.org/10.69777/349150}{https://doi.org/10.69777/349150}), and the Natural Sciences and Engineering Research Council of Canada (NSERC), scholarship number CGRS D -- 610229 -- 2026.

\appendix \addtocontents{toc}{\protect\setcounter{tocdepth}{1}}

	\section{One-loop graviton scattering}\label{sec:1loopApp}
	\subsection{Graviton scattering in 3d \texorpdfstring{$\mathcal{N}=8$}{N=8}}
    In this appendix, we consider the $\langle 2222\rangle$ Mellin amplitude in 3d $\mathcal{N}=8$ ABJM theory~\cite{Aharony:2008ug,Aharony:2008gk}, which is holographically dual to M-theory on AdS$_4\times S^7/\mathbb{Z}_k$ for $k=1,2$. In the large $c_T$ expansion, the Mellin amplitude is organized as:
    \begin{equation}
        \begin{aligned}
            \mathcal{M}(s,t;\sigma,\tau)&=c_T^{-1}M^R+c_T^{-\frac53}B^{R^4}_4M^4+c_T^{-2}(M^{R|R}+B^{R|R}_4M^4)\\
            &+c_T^{-\frac73}(B^{D^6R^4}_4M^4+B^{D^6R^4}_6M^6+B^{D^6R^4}_7M^7)\\
            &+c_T^{-\frac{23}{9}}(B^{D^8R^4}_4M^4+B^{D^8R^4}_6M^6+B^{D^8R^4}_7M^7+B^{D^8R^4}_8M^8)\\
            &+c_T^{-\frac83}(M^{R|R^4}+B^{R|R^4}_4M^4+B^{R|R^4}_6M^6+B^{R|R^4}_7M^7+B^{R|R^4}_8M^8)+\dots\;,
        \end{aligned}
    \end{equation}
where the $M$'s are functions of $s,t,\sigma,\tau$ with numerical coefficients $B$ which can depend on $k$, and $M^n$ denotes the degree-$n$ polynomial solution of the SCWI. Here $c_T$ is normalized as in~\cite{Alday:2021ymb}, so that $c_T=16$ for the free $\mathcal{N}=8$ theory of eight real scalars and eight Majorana fermions, and $\lambda^2_{\text{Stress}}=256/c_T$; this $c_T$ is $\frac23$ of the $c_T$ defined in (\ref{charges}). The tree-level $R$ and $R^4$ terms were computed in~\cite{Zhou:2017zaw,Chester:2018aca}, and the absence of a $D^4R^4$ term at order $c_T^{-\frac{19}{9}}$ was shown in~\cite{Binder:2018yvd}. $M^R$ is an exchange term, while the $R^4$ term $B^{R^4}_4M^4$ is a contact term (i.e.\ a polynomial in $s,t$). The one-loop graviton scattering amplitudes, the $R|R$ and $R|R^4$ terms above and the $R^4|R^4$ term at order $c_T^{-\frac{10}{3}}$ (which also receives an $R|D^6R^4$ contribution that we do not consider), were already computed in~\cite{Alday:2021ymb,Alday:2022rly}. In both the tree-level and one-loop cases, the computations were done for the full amplitude. In reduced form, the polynomial solutions $M^n$ were given in~\cite{Virally:2025nnl}, together with the part of the tree-level amplitude $M^R$ that is rational in $s,t$ once the sums over poles are performed; for the rest of $M^R$, a reduced form could only be found there in certain limits. The reduced one-loop amplitudes were not known.

    We begin with the tree-level $M^R$ amplitude, which is simpler than the one-loop case. We find that it is given in a particularly convenient form in~\cite{Alday:2020dtb}; with the sign and normalization of~\cite{Alday:2021ymb}, it reads:
    \begin{equation}
        \begin{aligned}
            &M^R(s,t;\sigma,\tau) = M^R_s(s,t;\sigma,\tau)+\tau^2M^R_s\left(t,s;\tfrac{\sigma}{\tau},\tfrac{1}{\tau}\right)+\sigma^2M^R_s\left(u,t;\tfrac{1}{\sigma},\tfrac{\tau}{\sigma}\right)\;,\qquad u=4-s-t\;,\\
            &M^R_s(s,t;\sigma,\tau) = \frac{64}{\pi^{3/2}}\sum_{m=0}^\infty \frac{(t-2)(u-2)+(s+2)((t-2)\sigma+(u-2)\tau)}{\Gamma(\frac{1}{2}-m)^2m!\Gamma(m+\frac{5}{2})(s-1-2m)}\;.
        \end{aligned}
    \end{equation}
    The poles of this amplitude reproduce the position space stress tensor superconformal block. In (\ref{eq:3d_max_reduced_blocks}), we found that there exists a single conformal block which gives the stress tensor exchange's contribution. Using this as an intermediary, the reduced Mellin transform (\ref{eq:3d_max_reduced_mellin}) allows us to find a reduced expression $\widetilde{M}^R$ which reproduces $M^R$ when acted upon by $\mathcal{D}_{3}$ (normalized as in~\cite{Virally:2025nnl}):
    \begin{equation}
        \begin{aligned}
        \widetilde{M}^R(s,t) ={}& \frac{1}{2}\Big[\widetilde{M}^R_s(s,t) + \widetilde{M}^R_s(t,s){}+\widetilde{M}^R_s(s,\tilde u)+\widetilde{M}^R_s(\tilde u,s)+\widetilde{M}^R_s(t,\tilde u)+\widetilde{M}^R_s(\tilde u,t)\Big]\\
        &+ \frac{32\big(16(s+t+1){}+16st+4(s^2+t^2)+3(s^2t+st^2)\big)}{3\pi^2 s(s+2)t(t+2)(s+t+2)(s+t+4)}\;,\qquad \tilde u=-4-s-t\;,\\
        \widetilde{M}^R_s(s,t) ={}& \frac{32\,\Gamma(\frac{1-s}{2})\Gamma(2+\frac{s}{2})\Gamma(\frac{3+s+t}{2})}{9\pi^{5/2}\Gamma(1-\frac{s}{2})\Gamma(\frac{5+s}{2})\Gamma(3+\frac{s+t}{2})}\, {}_3F_2\left(1-\frac{t}{2},\frac{1}{2},\frac{1}{2};\frac{5}{2},\frac{5+s}{2};1\right)\;.
        \end{aligned}
    \end{equation}
    Since $\mathcal{D}_3$ has a kernel, $\widetilde{M}^R$ is defined only up to elements of this kernel: for instance, $\widetilde{M}^R_s(s,t)$ and $\widetilde{M}^R_s(s,\tilde u)$ have the same image under $\mathcal{D}_3$. The expression above is a representative which obeys (\ref{eq:mellin_crossing_red_max}). It has poles at $s,t,\tilde u=0,-1,-2,-4,-6,\ldots$, whereas $M^R$ has poles at $s,t,u=1,3,5,\ldots$ only.

    We now consider the one-loop amplitudes. We begin with the simplest cases, which are the $R|R^4$ and $R^4|R^4$ amplitudes. These take the general form:
    \begin{equation}
        \begin{aligned}
        M^{A|R^4}(s,t;\sigma,\tau) ={}& \sum_{m=1,2,\ldots} \frac{d^{A|R^4}_{\text{even}}(s,t;\sigma,\tau;m)}{s-2m} + \delta_{k,1}\sum_{m=\frac{3}{2},\frac{5}{2},\ldots} \frac{d^{A|R^4}_{\text{odd}}(s,t;\sigma,\tau;m)}{s-2m}\\
        &+ \text{crossed} + P(s,t;\sigma,\tau)\;,
        \end{aligned}
    \end{equation}
    where $A=R$ or $R^4$, and the $d$ are quadratic in $\sigma,\tau$. The sums over integer and half-integer $m$ come from double-trace operators of even and odd twist, respectively; the latter are built from the graviton KK modes with odd $p$, which are projected out by the $\mathbb{Z}_2$ orbifold for $k=2$~\cite{Alday:2021ymb,Alday:2022rly}, hence the factor $\delta_{k,1}$. $P(s,t;\sigma,\tau)$ is an additional contact term, which is needed to solve the SCWI, and ``crossed'' denotes terms related to the explicit ones by crossing symmetry. We write the reduced amplitude as:
    \begin{equation}
        \widetilde{M}^{A|R^4}(s,t) = \sum_{m=1,2,\ldots} \frac{\tilde d^{A|R^4}_{\text{even}}(m)}{s-2m} + {\delta_{k,1}}\sum_{m=\frac{3}{2},\frac{5}{2},\ldots} \frac{\tilde d^{A|R^4}_{\text{odd}}(m)}{s-2m} + \text{crossed}\;,
    \end{equation}
    where we find that the reduced $\tilde{d}$ coefficients have no dependence on $s,t$. Here, there is no need for a contact contribution, as the action of $\mathcal{D}_{3}$ is sufficient to ensure the SCWI are obeyed. Explicitly, the $\tilde d$ coefficients are:
    \begin{equation}
        \begin{aligned}
            \tilde d^{R^4|R^4}_{\text{even}}(m) ={}& -\frac{143360\cdot{2^{5/3}}\,(64 m^5-288 m^4+624 m^3-744m^2+470 m-123) \Gamma (m+2)}{\sqrt[3]{3}\,\pi^{16/3}{k^{4/3}}\left(-\frac{3}{2}\right)_{m+1}}\;,\\
            \tilde d^{R^4|R^4}_{\text{odd}}(m) ={}&-\frac{1146880\cdot 2^{2/3} (64 m^5-288 m^4+624 m^3-744m^2+470 m-123) \Gamma (m+2)}{3 \sqrt[3]{3}\pi ^{29/6} k^{4/3}\Gamma \left(m-\frac{1}{2}\right)}\;,\\
            \tilde d^{R|R^4}_{\text{even}}(m) ={}&\frac{2048 \sqrt[3]{2} \left(-3 (5 m+4)-\frac{32\sqrt{\pi } (10 m (10 m-7)+27) \Gamma(m+2)}{\Gamma\left(m-\frac{1}{2}\right)}\right)}{105\cdot 3^{2/3}\pi ^{14/3} k^{2/3}}\;,\\
            \tilde d^{R|R^4}_{\text{odd}}(m)={}&-\frac{65536 \sqrt[3]{2} (100 m^2-70 m+27) \Gamma(m+2)}{105\cdot 3^{2/3} \pi ^{25/6} k^{2/3} \Gamma\left(m-\frac{1}{2}\right)}\;.
        \end{aligned}
    \end{equation}
    Finally, the $R|R$ case is the most complex one. The full amplitude is given by:
    \begin{equation}
        \begin{aligned}
        M^{R|R}(s,t;\sigma,\tau) ={}&\sum_{m,n=1,2,\ldots} \frac{c_{\text{even}}(s,t;\sigma,\tau;m,n)}{(s-2m)(t-2n)} + \delta_{k,1}\sum_{m,n=\frac{3}{2},\frac{5}{2},\ldots} \frac{c_{\text{odd}}(s,t;\sigma,\tau;m,n)}{(s-2m)(t-2n)}\\
        &+\sum_{m=1,2,\ldots} \frac{d^{R|R}_{\text{even}}(s,t;\sigma,\tau;{m})}{(s-2m)}+\delta_{k,1}\sum_{m=\frac{3}{2},\frac{5}{2},\ldots} \frac{d^{R|R}_{\text{odd}}(s,t;\sigma,\tau;m)}{(s-2m)}\\
        &+ \text{crossed} + P(s,t;\sigma,\tau)\;,
        \end{aligned}
    \end{equation}
    where the $c$ and $d$ coefficients have a known, but lengthy, form, and $P(s,t;\sigma,\tau)$ is once again a polynomial added to solve the SCWI. In the reduced amplitude, we find that we do not need the single pole contributions, and only need the double poles, exactly as we found for the gluon amplitudes in Section \ref{sec:1loop}:
    \begin{equation}
        \widetilde{M}^{R|R}(s,t) = \sum_{m,n=1,2,\ldots} \frac{\tilde c_{\text{even}}(m,n)}{(s-2m)(t-2n)} + \delta_{k,1}\sum_{m,n=\frac{3}{2},\frac{5}{2},\ldots} \frac{\tilde c_{\text{odd}}(m,n)}{(s-2m)(t-2n)} + \text{crossed}\;.
    \end{equation}
    Here ``crossed'' denotes the same two sums with $(s,t)$ replaced by $(s,\tilde u)$ and by $(t,\tilde u)$. As in the previous cases, the coefficients above have no dependence on $s,t$, and we do not need any additional contact terms, as $\mathcal{D}_3$ ensures the amplitude solves the SCWI. The sums in $\widetilde{M}^{A|R^4}$ and $\widetilde{M}^{R|R}$ diverge, however, and these reduced amplitudes are defined only up to crossing-symmetric polynomials in $s,t$, which $\mathcal{D}_3$ maps to contact terms of the kind that appear in the expansion of $\mathcal{M}$ above. Explicitly, the $\tilde c$ coefficients are:
    \begin{equation}
        \begin{aligned}
        \tilde c_{\text{even}}(m,n) ={}& \frac{16\Gamma(m+2)\Gamma(n+2)\Gamma\left(m+n-\frac{3}{2}\right)}{9\pi^{7/2}\Gamma\left(m+\frac{3}{2}\right)\Gamma\left(n+\frac{3}{2}\right)\Gamma(m+n+3)}\Bigg(96 m^2 n^2 (m+n)+304 m^2 n^2\\
        &-20 (m+n)^3-16 m n(m+n)^2-56 (m+n)^2+120 m n (m+n)\\
        &-27 (m+n)+80 m n+10\Bigg)\;,\\
        \tilde{c}_{\text{odd}}(m,n) ={}& \frac{16\Gamma(m)\Gamma(n)\Gamma\left(m+n-\frac{3}{2}\right)}{9\pi^{7/2}\Gamma\left(m-\frac{1}{2}\right)\Gamma\left(n-\frac{1}{2}\right)\Gamma(m+n+3)}\Bigg(96 m^3 n^2+80 m^3 n-12 m^3\\
        &+96 m^2 n^3+464 m^2 n^2+436m^2 n+84 m^2+80 m n^3+436 m n^2+400 m n\\
        &-69 m-12n^3+84 n^2-69 n-45\Bigg)\;.
        \end{aligned}
    \end{equation}

    \subsection{Graviton scattering in 6d \texorpdfstring{$\mathcal{N}=(2,0)$}{N=(2,0)}}
    The one-loop graviton scattering amplitude in 6d $\mathcal{N}=(2,0)$ was already computed in~\cite{Alday:2020tgi}. Here, we again want to simplify the answer somewhat. Unlike in the 3d case, a reduced amplitude was already known in 6d $\mathcal{N}=(2,0)$. However, the choice of Mellin transform in previous work made the reduced amplitude more complex than necessary.

    It was already observed in~\cite{Virally:2025nnl} that the reduced amplitudes in Mellin space previously used in the literature introduce additional poles in the reduced amplitude, compared to a minimal choice for $\mathcal{D}_6$. This means that, for example, contact diagrams have poles at $s,t,\tilde u=6$. Here, we will show that using the minimal-pole choice for $\mathcal{D}_6$ simplifies the one-loop amplitude. Our first step will be to define $\Upsilon_6$ in this context%
    \footnote{Compared to~\cite{Alday:2020tgi}, we define $\Upsilon_6$ with an extra factor of $\frac{1}{U}$, i.e.\ $\Upsilon_{6,\text{here}} = \Upsilon_{\text{there}}\frac{1}{U}$, which makes the superconformal block decomposition simpler.}:
    \begin{equation}
        \Upsilon_6= (z\bar z)^4 \nabla_6(z\bar z)^{-5}[(1-z\alpha)(1-\bar z\alpha)(1-z\bar\alpha)(1-\bar z\bar\alpha)]\;,
    \end{equation}
    which is defined such that:
    \begin{equation}
        \mathcal{G}(U,V;\sigma,\tau) = U^2\nabla_6\left[\frac{(1-z\alpha)(1-z\bar\alpha) h(z) - (1-\bar z\alpha)(1-\bar z\bar\alpha)h(\bar z)}{z-\bar z}\right]+\Upsilon_6\left[\widetilde{\mathcal{G}}(U,V)\right]\;,
    \end{equation}
    where $\nabla_6$ is the operator defined below (\ref{wardSolForm}), and we separated out the chiral sector, which is described by the holomorphic function $h(z)$ (the function $h_2(z)$ of~\cite{Alday:2020tgi}). Under this definition, the reduced correlator admits the block decomposition~\cite{Beem:2015aoa,Alday:2020tgi}, with the OPE coefficients normalized as in~\cite{Alday:2020tgi}:
    \begin{equation}
        \begin{aligned}
            \widetilde{\mathcal{G}}(U,V) ={}&2\lambda^2_{\mathcal{D}[40]}G_{8,0}^{0,-2}(U,V) + \frac{\lambda^2_{\mathcal{D}[04]}}{3}G_{10,0}^{0,-2}(U,V) \\
            &+ \sum_{\ell=0,2,\ldots} \frac{2(\ell+3)\lambda^2_{\mathcal{B}[20]_\ell}}{(\ell+5)(\ell+1)}G_{10+\ell,\ell+2}^{0,-2}(U,V)\\
            &+ \sum_{\ell=1,3,\ldots} \frac{(\ell+2)\lambda^2_{\mathcal{B}[02]_\ell}}{(\ell+1)(\ell+4)}G_{11+\ell,\ell+1}^{0,-2}(U,V)\\
            &-\sum_{\Delta}\sum_{\ell=0,2,\ldots}\frac{4\lambda^2_{\Delta,\ell}}{(6+\ell-\Delta)(\Delta+\ell-2)}G_{\Delta+4,\ell}^{0,-2}(U,V)\;.
        \end{aligned}
    \end{equation}
    The $\mathcal{D}[00]$ and $\mathcal{D}[20]$ blocks appear only in the chiral sector. We choose to define our Mellin transform for $\widetilde{\mathcal{G}}(U,V)$ as:
    \begin{equation}
        \begin{aligned}
            \widetilde{\mathcal{G}}(U,V) ={}& {-}32\int\frac{\text{d}s\,\text{d}t}{(4\pi i)^2}U^{\frac{s}{2}+2}V^{\frac{t}{2}-3}\Gamma\left(3-\frac{s}{2}\right)\Gamma\left(4-\frac{s}{2}\right)\Gamma\left(3-\frac{t}{2}\right)\Gamma\left(4-\frac{t}{2}\right)\\
            &\times\Gamma\left(3-\frac{\tilde u}{2}\right)\Gamma\left(4-\frac{\tilde u}{2}\right)\widetilde{\mathcal{M}}(s,t)\;,
        \end{aligned}
    \end{equation}
    where $s+t+\tilde u=10$. This Mellin transform is chosen such that we obtain $\mathcal{D}_6$ as given in~\cite{Virally:2025nnl}. Since $\Upsilon_6$ differs from the operator of~\cite{Alday:2020tgi} only by the factor of $\frac{1}{U}$, the Mellin transform used there for the reduced correlator differs from ours only in the gamma functions and in the normalization, and the two reduced Mellin amplitudes are related by
    \begin{equation}
        \widetilde{\mathcal{M}}(s,t) = \frac{(s-6)(t-6)(\tilde u-6)}{256}\,\widetilde{\mathcal{M}}_{\text{there}}(s,t)\;,
    \end{equation}
    where $\widetilde{\mathcal{M}}_{\text{there}}$ is the reduced Mellin amplitude of~\cite{Alday:2020tgi}. In particular, the $R^4$ contact term, which is proportional to $\frac{1}{(s-6)(t-6)(\tilde u-6)}$ in~\cite{Alday:2020tgi}, becomes a constant, and the poles at $s,t,\tilde u=6$ of the one-loop amplitudes of~\cite{Alday:2020tgi} are absent from $\widetilde{\mathcal{M}}$. In the language of~\cite{Alday:2020tgi}, the one-loop amplitude in 6d $\mathcal{N}=(2,0)$ is given by:
    \begin{equation}
        \widetilde{M}^{R|R}_{\text{there}}(s,t)=\sum_{m,n=4}^{\infty}\frac{c_{mn}}{(s-2m)(t-2n)}+\sum_{m=4}^{\infty}\frac{c_m}{s-2m}\left(\frac{1}{t-6}+\frac{1}{\tilde u-6}\right)+\text{crossed}\;,
    \end{equation}
    where the $c_{mn}=c_{nm}$ and $c_m$ are coefficients given in~\cite{Alday:2020tgi}, which differ between the $(2,0)$ theories of type $A_{N-1}$ and $D_N$, and ``crossed'' denotes the first sum with $(s,t)$ replaced by $(s,\tilde u)$ and by $(t,\tilde u)$, and the second sum with $s$ exchanged with $t$ and with $\tilde u$. In our language, however, only the first sum is needed, and we can write (up to a constant):
    \begin{equation}
        \widetilde{M}^{R|R}(s,t)=\sum_{m,n=4}^{\infty}\frac{\tilde c_{mn}}{(s-2m)(t-2n)}+\text{crossed}\;,
    \end{equation}
    where ``crossed'' is as for the first sum above, and the $\tilde c_{mn}$ are given by:

    \begin{equation}
        \tilde c_{mn}=-\frac{(m-3)(n-3)(m+n-2)}{32}\,c_{mn}\;.
    \end{equation}
    The second sum drops out because the new factor of $(s-6)(t-6)(\tilde u-6)$ kills off its poles at $s,t,\tilde u=6$, which are now captured by the gamma functions of the reduced Mellin transform. The remaining single poles are then killed due to the identity:
    \begin{equation}
        c_m = -\sum_{n\ge 4}\frac{m+2n-5}{m+1}c_{mn}\;.
    \end{equation}

    \section{Tree-level correlators in position space}\label{sec:tree_position}
    In this appendix, we give details of the transformation of the tree-level correlators into position space, which we need to obtain the tree-level CFT data by decomposing the position space correlators in conformal blocks. In order to perform this decomposition, we want expansions near $V=1$, but (\ref{eq:gluon_tree_mellin}) naturally gives an expansion near $V=0$. In order to obtain the correct expansion, we must first resum the expansion near $V=0$ and re-expand. In 6d, it is simple to do this by writing the amplitudes in terms of $\overline{D}_{r_1,r_2,r_3,r_4}(U,V)$ functions%
    \footnote{We use $\overline{D}_{r_1,r_2,r_3,r_4}(U,V)=\int\frac{\text{d}s\,\text{d}t}{(4\pi i)^2}\,U^{\frac{s}{2}}V^{\frac{t}{2}}\,\Gamma\bigl(-\frac{s}{2}\bigr)\Gamma\bigl(-\frac{t}{2}\bigr)\Gamma\bigl(r_2+\frac{s+t}{2}\bigr)$\\{}$\times\,\Gamma\bigl(\frac{r_3+r_4-r_1-r_2-s}{2}\bigr)\Gamma\bigl(\frac{r_1+r_4-r_2-r_3-t}{2}\bigr)\Gamma\bigl(\frac{r_1+r_2+r_3-r_4+s+t}{2}\bigr)$, as in~\cite{Alday:2021ymb}.}:
    \begin{equation}
        \begin{aligned}
            \mathcal{G}^{(6d)}_s(U,V;\alpha) = {}& \frac{U^2}{\Gamma(2p-2)}\Big(2p(1-\alpha)\overline{D}_{3,3,2p+1,2p+1}(U,V) + p(2-4\alpha)\overline{D}_{4,2,2p+1,2p+1}(U,V)\\
            &-2p\alpha\overline{D}_{4,2,2p,2p+2}(U,V)+(4p+3)(\alpha-1)\overline{D}_{3,2,2p+1,2p}(U,V)\\
            &+3\alpha(2p+1)\overline{D}_{3,2,2p,2p+1}(U,V)\Big)\;,\\
            \mathcal{G}^{(6d)}_t(U,V;\alpha) ={}&\frac{U^4}{\Gamma(2p-2)V^2}\Big(p(\alpha-1)^2\overline{D}_{4,2,2p,2p+2}(U,V)+p(\alpha^2-1)\overline{D}_{6,2,2p-2,2p+2}(U,V)\\
            &-p(2p+1)(\alpha^2-1)\overline{D}_{5,2,2p-2,2p+1}(U,V)\\
            &+(2p+1)(\alpha-1)(p\alpha-p+1)\overline{D}_{5,2,2p-1,2p}(U,V)-p(\alpha-1)^2\overline{D}_{6,2,2p,2p}(U,V)\\
            &-(2p+1)(\alpha-1)(p\alpha-p-\alpha)\overline{D}_{4,2,2p-1,2p+1}(U,V)\Big)\;,\\
            \mathcal{G}^{(6d)}_u(U,V;\alpha) ={}& \frac{U^4}{\Gamma(2p-2)}\Big({-}2p\alpha(\alpha-1)\overline{D}_{2,4,2p,2p+2}(U,V)\\
            &+(4p^2-1)\alpha(\alpha-1)\overline{D}_{2,4,2p-1,2p+1}(U,V)+(1-4p^2)\alpha\overline{D}_{3,4,2p-2,2p+1}(U,V)\\
            &-2p(\alpha-2)\alpha\overline{D}_{3,4,2p-1,2p+2}(U,V)+2p\alpha\overline{D}_{4,4,2p-2,2p+2}(U,V)\Big)\;.
        \end{aligned}
    \end{equation}
    Here $\mathcal{G}^{(6d)}_{a}(U,V;\alpha)$, with $a=s,t,u$, is the transform (\ref{eq:mellin_22pp}) of $\mathcal{M}_{p}^a(s,t;\alpha)$ for $d=6$, so that (\ref{eq:gluon_tree_mellin}) gives $\mathcal{G}_{F,p}^{(6)}=5\,\big(\mathbf{c}_s\mathcal{G}^{(6d)}_{s}+\mathbf{c}_t\mathcal{G}^{(6d)}_{t}+\mathbf{c}_u\mathcal{G}^{(6d)}_{u}\big)$. In all other cases, it is simplest to find an ansatz which performs the resummation, and compare terms of a power series near $V=0$. In 3d and 5d when $p$ is even, we only need the coefficient of $\log U$, since this gives us the anomalous dimensions. We find that the following ansatz gives us a resummation of the correlators:
    \begin{equation}
        \begin{aligned}
            \mathcal{G}_{F,p}^{(d)}(U,V;\alpha)\Big|_{\alpha^2\mathbf{c}_s\log U} = \sum_{n=0}^\infty\Big(&P_{n}^{(0)}(V) + P_{n}^{(1)}(V)\log V\\
            &+ P_{n}^{(2)}(V)\sqrt{1-V}\tanh^{-1}(\sqrt{1-V})\Big)U^{\frac{d-2}{2}p+n}\;,
        \end{aligned}
    \end{equation}
    where the $P_{n}^{(i)}(V)$ are polynomials in $V$ divided by a power of $V-1$. In 3d and 5d when $p$ is odd, $[\mathcal{O}_p\mathcal{O}_p]_{n,\ell}$ has odd twist and is absent from the generalized free $\mathcal{O}_2\times\mathcal{O}_2$ OPE, so there is no $\log U$ term at this order. Instead, we compute the half-integer powers of $U$ in the $\alpha^2\mathbf{c}_s$ component, using the ansatz of~\cite{Alday:2022rly}
    \begin{equation}
        \begin{aligned}
            \mathcal{G}_{F,p}^{(d)}(U,V;\alpha)\Big|_{\alpha^2\mathbf{c}_s,\;U^{\mathbb{Z}+\frac12}} = \sum_{n=0}^\infty\Big(&P_{n}^{(0)}(V) + P_{n}^{(1)}(V)K(1-V)\\
            &+ P_{n}^{(2)}(V)E(1-V)\Big)U^{\frac{d-2}{2}p+n}\;,
        \end{aligned}
    \end{equation}
    where $K(m)=\frac{\pi}{2}\,{}_2F_1\bigl(\frac12,\frac12;1;m\bigr)$ and $E(m)=\frac{\pi}{2}\,{}_2F_1\bigl(-\frac12,\frac12;1;m\bigr)$ are the complete elliptic integrals of the first and second kind, and once more the $P_{n}^{(i)}(V)$ are polynomials in $V$ divided by a power of $V-1$.

\bibliographystyle{JHEP}
\bibliography{Gloop}

\end{document}